%% file: article_last_arxiv.tex
\documentclass[12pt,preprint]{aastex}
\usepackage{apjfonts}
\usepackage{graphicx}
\usepackage{psfig}
\usepackage{amssymb}
\newcommand{\begit}{\begin{itemize}}
\newcommand{\enit}{\end{itemize}}
\newcommand{\begen}{\begin{enumerate}}
\newcommand{\enen}{\end{enumerate}}

\newcommand       \be           {\begin{equation}}
\newcommand       \ee           {\end{equation}}
\newcommand       \bea          {\begin{eqnarray}}
\newcommand       \eea          {\end{eqnarray}}

\newcommand       \cm		{\,{\rm cm }}

\newcommand       \yr		{\,{\rm yr }}

\newcommand       \s		{\,{\rm s }}

\newcommand       \La		{\,{\cal L }}
\newcommand       \C		{\,{\cal C }}
\newcommand       \G		{\,{\cal G }}
\newcommand       \Ha		{\,{\cal H }}
\newcommand       \kpc		{\,{\rm kpc }}

\newcommand       \ic           {\imath_{crit}}
\newcommand{\beqa}{\begin{eqnarray}} 
\newcommand{\eeqa}{\end{eqnarray}}

\def\mpy{\rm \ M_\odot \ {\rm yr^{-1}}}

\shorttitle{ON THE DYNAMICS OF 4U1820-30}
\shortauthors{Prodan \& Murray}

\begin{document}

\title{ON THE DYNAMICS AND TIDAL DISSIPATION RATE OF
  THE WHITE DWARF IN 4U 1820 -30}

\author{Sne\v zana Prodan\altaffilmark{1} \& Norman Murray\altaffilmark{1,2}}

\altaffiltext{1}{Canadian Institute for Theoretical Astrophysics, 60
St.~George Street, University of Toronto, Toronto, ON M5S 3H8, Canada;
sprodan@cita.utoronto.ca} 
\altaffiltext{2}{Canada Research Chair in Astrophysics}

\begin{abstract}
It has been suggested that the 170 day period in the light curve of
the low mass X-ray binary 4U 1820-30 arises from the presence of a
third body with a large inclination to the binary orbit. We show that
this long period motion arises if the system is librating around the
stable fixed point in a Kozai resonance.  We demonstrate that mass
transfer drives the system toward this fixed point, and calculate,
both analytically and via numerical integrations, that the period of
libration is of order 170 days when the mutual inclination is near the
Kozai critical value.  The non-zero eccentricity of the
binary, combined with tidal dissipation, implies that the rate of
change of the binary period would be slower than, or even of opposite
sign to, that implied by standard mass transfer models. If the 170 day
period results from libration, then, contrary to appearances, the
orbital period of the inner binary is increasing with time; in that
case, $(e/0.009)^2Q/k_2 \gtrsim 2.5\times10^9$, where $k_2\approx0.01$
is the tidal Love number and $e=0.009$ is the fiducial eccentricity of
the inner binary. It appears unlikely that the observed negative
period derivative results from the smaller than expected (but
positive) value of $\dot P$ combined with the previously suggested
acceleration of the system in the gravitational field of the host
globular cluster NGC 6624.  The discrepancy between the observed
and expected period derivative requires further investigation.
\end{abstract}


\section{INTRODUCTION}
4U 1820-30 is a low mass X-ray binary (LMXB) located near the
center of the globular cluster NGC 6624. The binary orbital period is
$P_{1}\simeq685\s$, revealed in X-ray
observations as a modulation with $\sim 2-3\%$ peak to peak amplitude \citep{1987ApJ...312L..17S}. Subsequently,
\citet{1997ApJ...482L..69A} discovered a $\sim16\% $ peak to
peak modulation (period $ 687.6\pm2.4$ s) in the UV band from HST.

This short period, low amplitude variation is very stable, with
$\dot{P}/P=(-3.47\pm1.48)\times10^{-8}yr^{-1}$
\citep{2001ApJ...563..934C}, which is consistent with the earlier
measurement of $\dot{P}/P=(-5.3\pm1.1)\times10^{-8}yr^{-1}$ from
\citet{1993A&A...279L..21V}; this
stability led \citet{2001ApJ...563..934C} to suggestion that this
modulation reflects the orbital period of the binary.

Both the short binary period and the type I X-ray bursts observed in
this system imply that the secondary star is a helium white dwarf, of mass
$m_2=(0.05-0.08) M_{\bigodot}$, accreting mass onto a primary neutron
star \citep{1987ApJ...322..842R}. The distance to the source is
estimated to be $7.6\pm0.4 \kpc$ \citep{2003A&A...399..663K}.

It is striking that neither the magnitude nor the sign of the period
derivative is consistent with the prediction
$\dot{P}/P>+8.8\times10^{-8}yr^{-1}$ of the standard evolution
scenario for compact binaries overflowing their Roche lobe
\citep{1987ApJ...322..842R}. It has been suggested that the negative
period derivative is only apparent, i.e., that it is not intrinsic to
the binary, but instead reflects the acceleration of the binary in the
gravitational potential of the globular cluster which houses the binary
\citep{1993MNRAS.260..686V}. However, quantitative estimates
show that the acceleration, while of roughly the right magnitude, is
unlikely to be large enough, by itself, to explain the large
discrepancy between the evolution scenario and the observations
\citep{1993MNRAS.260..686V,1993ApJ...413L.117K, 2001ApJ...563..934C}.

A second striking property of 4U 1820-303 is the much larger
luminosity variation, by factor of $\gtrsim2$, seen at a period of $P_{3}\simeq 171$ days.  Analysis of the RXTE ASM data shows
that this long period modulation does not exhibit a significant period
derivative, $\dot P_3/P_3<2.2\times10^{-4}yr^{-1}$\citep{2001ApJ...563..934C}. The ratio between this long
period and the binary orbital period is $\simeq 2\times10^4$, which
appears to be too high to be due to disk precession at the mass ratio
of the system \citep{1998MNRAS.299L..32L, 1999MNRAS.308..207W}.

In this paper we adopt the assumption of
\citet{1988IAUS..126..347G}, that the $171$ day period is due to the
presence of a third body in the system. The third (outer) star
modulates the eccentricity of the binary at long term period
$P_{3}\simeq P^{2}_{2}/(eP_{1})$, where $P_{2}$ is the orbital period of
the third star and $e$ is the eccentricity of the inner binary. Taking into account only perturbations from the third star, the binary orbital period of $685~\s$ and $\sim171$ day
long-term modulation imply that the orbital period of the third star
must be $\sim 1$ day. The presence of additional sources of precession, such as that due to tidal distortion of the white dwarf secondary, requires a stronger perturbation from the third body and hence a smaller orbit in order to modulate eccentricity of the inner binary at the $171$ day period.  We show that the luminosity modulation arises
from variations in the eccentricity of the inner binary associated with
libration around a stable fixed point in the Kozai resonance.

Tidal dissipation in the white dwarf, driven by the eccentricity
of the binary orbit, tends to decrease both the eccentricity and the
semimajor axis (hence period) of the binary, which we suggest is
responsible, in part, for the anomalous observed period
derivative---note that \citet{1987ApJ...322..842R} did not treat the
effects of tidal dissipation. The combination of tidal dissipation and
mass transfer will result in a lower value of $\dot{P}/P$ than that
produced by conservative mass transfer alone.

For rapid enough dissipation, or, expressed another way, for low
enough values of the tidal dissipation parameter $Q$, $\dot P<0$ could
result. We do not favor this as the explanation for the observed
negative period derivative; we show that such rapid dissipation damps
eccentricity within $10^{-3}$ of the system's lifetime. Subsequently
the mass transfer takes over the evolution of the semimajor axis. In
other words, we would be incredibly lucky to observe the system in the
short time that $e$ is significant, in the absence of another
perturbing influence.  We also show that, given the most recent estimates for
the acceleration of millisecond pulsars in the
gravitational field of the globular cluster, the cluster gravity does not appear to
contribute significantly to the observed period derivative of 1820-30.

Thus it appears that, while both tidal dissipation and acceleration in the gravitational field of the cluster contribute negatively to the period derivative, they can not fully explain it. Since we favor the hierarchical triple model as an explanation for the origin of $171$day period of luminosity variations, we suggest that the apparent negative period derivative, which is a $2-\sigma$ result, may either be an observational artifact or due to the some yet not understood physical processes.

The relation between the luminosity variations and the period
derivative is deeper; we argue that the (intrinsic) increase in the
semimajor axis of the binary (driven by Roche lobe overflow) leads to
trapping of the system deep in the Kozai resonance. The resonance
transfers angular momentum from the inner binary to the third star,
and back, periodically, without affecting the semimajor axis of either
orbit. However, the dissipation associated with the strong tides when
the forced eccentricity is largest does remove energy from the orbit
of the inner binary. This energy loss peaks when the mutual
inclination is small. It is well known that this coupled Kozai-tidal
evolution tends to leave the system with a mutual inclination between
the two orbits near the Kozai critical value ($\sim40^\circ$); see,
for example, Figure 4 in \citet{2007ApJ...670..820W} or Figure 7 in
\citet{2007ApJ...669.1298F}. We show that the period of small
oscillations is naturally $\sim170$ days when the mutual inclination is close to the Kozai critical value. Whether the evolution of the inclination in systems like 1820-30, which, unlike the planetary systems, is know to undergo Roche lobe overflow, is a question we are currently investigating.

This paper is organized as follows. In \S \ref{sec:dynamics} we
develop an analytic understanding of the system, describing the
resonance dynamics, calculating the location of the fixed point as a
function of the system parameters (stellar masses, orbital radii, and
the mutual inclination of the two orbits), and the frequency (or
period) of small oscillations. In \S \ref{sec:capture} we describe a
possible dynamical path by which the system arrived at its present
configuration. The dynamical history relies crucially on both the
Roche lobe overflow (which drives the system into resonance) and the
tidal dissipation, which tends to drive the mutual inclination toward
the Kozai critical value. In section \ref{sec:numerics} we describe
the results of numerical integrations of the equations of motion,
presenting a fiducial model that reproduces the observed properties of
4U 1820-30. We also demonstrate trapping in the case of an expanding
inner binary orbit, and detrapping in the case of a shrinking binary
orbit. In \S \ref{sec:constraints} we use the model to put constraints
on the ratio of the tidal dissipation parameter
$Q$ and the tidal Love number ($k_2$) of the Helium white dwarf for our fiducial eccentricity. We discuss our results, and those of previous workers,
in \S \ref{sec:discussion}. We present our conclusion in the final
section. We give the details of the numerical model in the appendix A. In appendix B we discuss in details adiabatic invariance of the action and how it governs the evolution of the system by comparing analytic and numerical analysis.

\section{UNDERSTANDING THE DYNAMICS OF THE $4U 1820-30$ SYSTEM} \label{sec:dynamics}
The presence of a third body orbiting the center of mass of a tight
binary will induce changes in the orbital elements of the binary,
changes that take place over a variety of time scales. The changes are
particularly dramatic if the mutual inclination of the two orbits is
large. \citet{1962AJ.....67..591K} showed that when the
initial inclination between inner and outer orbits has values between
some critical inclination $\ic$ and $180^{o}-\ic$, both the
eccentricity of the inner binary and the mutual inclination undergo
periodic oscillations known as Kozai cycles.

The period of the Kozai cycles is much longer than either the binary's
orbital period, or the period of the outer orbit. This justifies the
use of the secular approximation, which involves averaging the
equations of motion over the orbital periods of inner and outer
binaries; as a result, the averaged equations of motion predict that
the semimajor axes of both binaries are unchanged.

If the luminosity variations in 4U 1820-30 are due to Kozai cycles,
the semimajor-axis ratio $a_{out}/a \approx 8$, so in our analytic work we use the
quadrupole approximation for the potential experienced by the inner
binary due to the third body. In our numerical work we keep terms to octupole order, but we show that the higher order terms change the quantitative results only slightly.

The angular momentum of the outer binary is much greater than that of
the inner, so that the orientation of the outer binary is, to a good
approximation, also a constant of the motion. In that case, after the
averaging procedure, the final Hamiltonian has one degree of freedom.

Kozai cycles are the consequence of a $1 : 1$ resonance between the
precession rates of the longitude of the ascending node $\Omega$ and
the longitude of the periastron $\varpi$ of the inner binary. The
condition for Kozai resonance, $\dot{\varpi}-\dot{\Omega}=0$, is
satisfied only for high inclination orbits; for low inclinations, the
line of nodes precesses in a retrograde sense ($\dot\Omega<0$), while
the apsidal line precesses in a prograde sense.

We employ Delaunay variables to describe the motion of the inner
binary. The angular variables are the mean anomaly $l$, the argument
of periastron $\omega$, and the longitude of the ascending node
$\Omega$; of these, only $\omega$ appears in the averaged
Hamiltonian. Their respective conjugate momenta are:
\bea \La &=&m_1
m_2 \sqrt{\frac{Ga}{m_1+m_2}}\\ \G &=&\La\sqrt{1-e^2}\\ \Ha
&=&\G\cos{\imath}.  \eea
The longitude of periastron is $\varpi\equiv\Omega+\omega$.  Recall
that we are assuming that the semimajor axis of the outer binary is
large enough that the total angular momentum is dominated by that of
the outer binary, so that $\imath$ is effectively the mutual
inclination between the two binary orbits. We occasionally refer to
the elements of the third star, using a subscript 'out' to distinguish
them from those of the inner binary.

After averaging over $l$ and $l_{out}$, the Hamiltonian describing the
motion of a tight binary orbited by a third body, allowing for the
effects of both tidal and rotational bulges on the secondary, and for
the apsidal precession induced by general relativistic effects, is
\citep{1997AJ....113.1915I,2000ApJ...535..385F,2007ApJ...669.1298F}
\bea \label{eq:hamiltonian} %$ 
H&=&\frac{-3A}{2}\Bigg[-{5\over3}-3\frac{\Ha^{2}}{\La^{2}}+\frac{\G^{2}}{\La^{2}}+5\frac{\Ha^{2}}{\G^{2}}+5\cos2\omega\left(1-\frac{\G^{2}}{\La^{2}}-\frac{\Ha^{2}}{\G^{2}}+\frac{\Ha^{2}}{\La^{2}}\right)\Bigg]\nonumber\\ &&-B\frac{\La}{\G}-k_2C\left(35\frac{\La^{9}}{\G^{9}}-30\frac{\La^{7}}{\G^{7}}+3\frac{\La^{5}}{\G^{5}}\right)
-k_2D\frac{\La^{3}}{\G^{3}}, 
\eea %$
where the term proportional to $A$ is the Kozai term, the term
proportional to $B$ enforces the average apsidal precession due to
general relativity, and the terms proportional to $C$ and $D$
represent the tidal and rotational bulges, respectively; the explicit
appearance of the tidal Love number $k_2$ in the latter two terms
highlights the fact that these terms represent the effects of the
white dwarf's tidal and rotational bulges. The expressions for the
constants are
\bea %$
A&=&{1\over8}\Phi\,
{m_2m_3\over(m_1+m_2)^2}
\left({a\over a_{out}}\right)^3
{1\over(1-e^2_{out})^{3/2}}\\ 
B&=&{3\over 2}\Phi
 {m_2\over m_1} {r_s\over a}\label{eq:GRa}\\
C&=&{1\over 16}\Phi
{m_1\over m_1+m_2}\left({R_2\over a}\right)^{5}\\
D&=&{1\over12}\Phi
\left({R_2\over a}\right)^5f(\tilde\Omega_{spin})\label{eq:spin},
\eea %$
where
\be  %$
\Phi\equiv{G(m_1+m_2)m_1\over a}.
\ee  %$
Recall that the semimajor axis and eccentricity of the outer body's
orbit are denoted by $a_{out}$ and $e_{out}$. The quantity
$r_s\equiv2Gm_1/c^2$ in equation (\ref{eq:GRa}) is the Schwarzschild
radius of the neutron star.

As just noted, the term proportional to $D$ accounts for the
rotational bulge produced by the spin of the white dwarf. The spin is
projected onto the triad defined by the Laplace-Runge-Lenz vector,
pointing along the apsidal line from the white dwarf at apoapse toward
the neutron star, and denoted by a subscript $e$, the total angular
momentum vector, subscript $h$, and their cross product, denoted by
$q$. We have scaled the spin to the orbital frequency (or mean motion)
$n$, so that, e.g., $\tilde\Omega_e\equiv\Omega_e/n$. We do so because
we anticipate that for small eccentricity the white dwarf will be
tidally locked. Then
$f(\tilde\Omega_{spin})\equiv2\tilde\Omega^2_h-\tilde\Omega_e^2-\tilde\Omega_q^2$
is a dimensionless quantity of order unity.

For the fiducial values of the system parameters listed in table 1, $A\approx1.73\times10^{44}$,
the ratios $B/A\approx 0.53$, $C/A\approx1.82$, and $D/A\approx2.54$.

\input{table1}

\subsection{The Kozai mechanism}
We start our discussion of the dynamics of the system by focusing on
understanding the Kozai mechanism, neglecting forces due to the tidal
and rotational bulges of the Helium white dwarf in the inner binary,
and the effects of general relativity.

We locate the resonance by looking for a fixed point of the
Hamiltonian; since we are neglecting the tidal and rotational bulges,
and the general relativistic precession, we set $B=C=D=0$ and
differentiate the Hamiltonian with respect to $\omega$, to find
$\omega_f=0,90^\circ,180^\circ,270^\circ$. The fixed points at $\omega_f=90^\circ$ and
$\omega_f=270^\circ$ are stable. Differentiating the Hamiltonian with
respect to $\G$, substituting $\omega=90^\circ$ (or $270^\circ$) and setting the
result equal to zero, we find $\G_f^4=(5/3)\Ha^2\La^2$. In terms of
the eccentricity,
\be \label{eq:fixed_e}%$
e_f = \sqrt{1-{5\over3}\cos^2\imath_f},
\ee %$
where the subscript $f$ indicates that this is the eccentricity of the
stable fixed point. The frequency of small oscillations around the
fixed point (small librations) is
\be %$ 
\omega_0\equiv \left[\left({\partial^2 H\over \partial
\omega^2}\right)_{\omega_f,\G_f}\left({\partial^2H\over\partial\G^2}\right)_{\omega_f,\G_f}\right]^{1/2}.
    \ee Performing the derivatives, \be \label{eq:omega_0} %$
    \omega_0=\omega_A\left(18+90{\Ha^2\La^2\over\G_f^4}\right)^{1/2}
    \left(1-{\G^2_f\over \La^2}-{\Ha^2\over\G_f^2}+{\Ha^2\over
      \La^2}\right)^{1/2}, \ee %$
where we have defined 
\be %$
\omega_A \equiv \sqrt{30A^2\over \La^2}=\sqrt{15\over32}n{m_3\over m_1+m_2}
\left({a\over a_{out}}\right)^3{1\over(1-e^2_{out})^{3/2}}.
\ee %$

The last factor in equation (\ref{eq:omega_0}) is $e_f\sin\imath_f$.

In terms of the eccentricity,
\be \label{eq:omega_0e}%$ 
\omega_0 = {3\over2}\sqrt{15}\,n{m_3\over
m_1+m_2}\left({a \over a_{out}}\right)^3{e_f\sin\imath_f\over
    (1-e_{out}^2)^{3/2}} . \ee %$

From equation (\ref{eq:fixed_e}) we see that the critical inclination for a
Kozai resonance to occur, in the absence of other dynamical effects,
is $\ic = \cos^{-1} \sqrt{3/5}\approx 39.2^\circ$. If $\imath>\ic$,
orbits started at $\omega=90^\circ$ with $e<e_f$ will librate around
the fixed point, so that $\omega$ remains between $0^\circ$ and
$180^\circ$ (or an even more restricted range). From
equation (\ref{eq:omega_0}) or (\ref{eq:omega_0e}), the period of small
oscillations $P_0\sim 1/e_f$, a point that will be important later.

In contrast, orbits started at $\omega=0$ and $e>0$ will circulate
($\omega$ will range from $0$ to $360^\circ$). Librating and
circulating orbits are separated by the separatrix, an orbit that
neither librates nor circulates. The width of the separatrix (as
measured by the excursion in $e$) depends only on the initial
inclination: $e_{sep}=[1-(5/3)\cos^{2} \imath]^{1/2}$.

Examples of librating and circulating orbits (for a system including
the effects of GR and tidal bulges) are shown in section 
\ref{sec:numerics}.

Note that, even for systems with $\imath<\ic$, where no stable Kozai
fixed point exists, both the mutual inclination and the eccentricity
of the inner binary can undergo oscillations with significant
amplitude (although reduced compared to the case with $\imath>\ic$).

Kozai cycles will be substantial only as long as the perturbation from
the outer body dominates over the other sources of apsidal precession
in the inner binary orbit, a point we now address.

\subsection{Kozai cycles in the presence of additional forces}

The physical effects represented by the terms proportional to $B$,
$C$, and $D$ are capable of suppressing Kozai oscillations. We
investigate their effects in this section.

As an aside, there is a small apsidal precession introduced by
dissipative effects in the He white dwarf, but this precession rate is
negligible compared to the other three. We mention it here because
tidal dissipation has a major role to play in the capture (or
otherwise) of the system into the Kozai resonance.

The equations for the precession rates due to the Kozai mechanism,
that due to general relativity, and the tidal and rotational bulges of
the white dwarf, are:
\pagebreak
\bea \label{eq: eqn for omega_dots}
\dot{\omega}_{Kozai}&=&{3\over4}n
\left({m_3\over m_1+m_2}\right)
\left({a\over a_{out}}\right)^3
{1\over(1-e_{out}^2)^{3/2}}\times
{1\over\sqrt{1-e^2}}\nonumber\\
&&\qquad\left[ 2(1-e^2)+5\sin^2\omega(e^2-\sin^{2}\imath)\right] \label{eq:kozai}\\
\dot{\omega}_{GR}&=&{3\over2}n\left({m_1+m_2\over m_1}\right)
\left({r_s\over a}\right)
{1\over(1-e^2)}\label{eq:GR}\\
\dot{\omega}_{TB}&=&{15\over16}n\,k_2{m_1\over m_2}
\left({R_2\over a}\right)^5{8+12e^2+e^4\over(1-e^2)^5}\label{eq:tb}\\
\dot{\omega}_{RB}&=&{n\,k_2\over4}{m_1+m_2\over m_2}
\left({R_2\over a}\right)^5{1\over(1-e^2)^2}
\Bigg[
\left(2\tilde\Omega^{2}_{h}-\tilde\Omega^{2}_{e}-\tilde\Omega^{2}_{q}\right)+2\tilde\Omega_{h}\cot\imath\left(\tilde\Omega_{e}\sin\omega +\tilde\Omega_{q}\cos\omega\right)
\Bigg].\label{eq:rb}
\eea

The Kozai term (equation \ref{eq:kozai}) can be either positive or
negative, depending on the value of $\sin \imath$. Both the white
dwarf tidal bulge and the GR terms are positive, so both tend to
suppress Kozai oscillations. The term induced by the white dwarf
rotational bulge, on the other hand, can be of either sign, depending
on the orientation of the white dwarf spin. If the white dwarf is
tidally locked and if its spin is aligned (which we assume in our
analytic model, but not in our numerical models), this term
contributes positive $\dot{\omega}$. In case of non-aligned spins the
precession rate may be negative (as we will see).

\subsubsection{The tidal bulge and the tidal Love number $k_2$}
The tidal bulge of the white dwarf in 4U 1820-30 dominates the non-Kozai
apsidal precession rate, for physically plausible values of $k_2$. We
remind the reader that in Newtonian gravitational theory the tidal
Love number $k_2$ is a dimensionless constant that relates the mass
multipole moment created by tidal forces on a spherical celestial body
to the gravitational tidal field in which it is immersed; in other
words, $k_2$ encodes information about body's internal
structure.\footnote{In a confusing usage, the apsidal precession
  constant, which is a factor of two smaller than the tidal Love
  number, but which we do not employ, is also denoted by $k_2$.}
  
  We use $k_2=0.01$, which is computed by Arras (private communication) as the ratio of the potential due to the perturbed mass distribution, to the external potential causing the perturbed mass, under the assumption that our He white dwarf is a fluid object.
  
  Soft X-ray observations of the source indicate a rather small absorption, consistent with that expected to be produced by the interstellar medium of the Galaxy; this rules out any significant outflows from the accretion disk or the surface of the white dwarf.  This implies an absence of mass loss through the $L_2$ Lagrangian point of the white dwarf, which puts an upper limit on the eccentricity of the inner binary; according to \citet{2005MNRAS.358..544R}, for our system parameters,  the upper limit on the eccentricity of inner binary is $e_{max}\simeq 0.07$.

If 4U 1820-30 has a non-zero but small eccentricity, as indicated by the
observed luminosity variations, then in the absence of a third body,
the precession rate of the binary orbit is dominated by the tidal
bulge induced in the white dwarf by the gravity of the neutron star;
from equations (\ref{eq:GR}) and (\ref{eq:tb}), the tidal bulge induces a
precession rate at least a few times that induced by GR:
\be  %$
{\dot\omega_{TB}\over\dot\omega_{GR}} \approx 4 
\left({k_2\over0.01}\right)
\left({a\over 1.32\times10^{10}\cm}\right)^{-4}.
\ee 

In order for the Kozai mechanism to produce significant variations in
$e$, the Kozai-induced precession rate must be comparable to or larger
than the sum of the precession rates produced by the other terms. For
physically realistic values of $k_2$, as we have just seen, the
precession rate induced by the tidal bulge of the white dwarf is by
far the largest, so if the Kozai effect is to be important, it must
produce a precession rate larger than $\dot \omega_{TB}$.

\subsection{Libration around the fixed point and the frequency of small oscillations}

\subsubsection{Why libration?}

For the values of the tidal Love number $k_2$ and eccentricity listed in table 1, the period of
the precession rate induced by the tidal bulge,
$P_{TB}=2\pi/\dot\omega_{\rm TB}$, is a factor of ten shorter that the
period of the observed luminosity variations. If this term set the
rate of precession, and the eccentricity varied as a result of this
precession, then the variations in X-ray luminosity would occur with a
period substantially shorter than the observed $170$ days.

In order to produce a much longer period, some other term must tend to
produce a negative precession rate. When this negative precession rate
is added to that produced by the tidal bulge, the resulting period can
be much longer than that produced by the tidal bulge alone.

Under the assumption that the white dwarf is tidally locked (we show later it is not), the only
term capable of producing a negative precession rate is the Kozai
term. Hence we are led to look for a cancellation between the Kozai
precession rate and the precession rate induced by the tidal bulge.

However, it is not enough to ask for a rough cancellation. To get the
observed precession rate, the sum of all the terms must cancel to
better than $10\%$. This requires some fine tuning of the mutual
inclination, a rather unsatisfactory situation.

On the other hand, if the system is captured into libration, then the
sum of all the precession terms is exactly zero. If the system is deep
in the resonance, then the period of libration is simply the period
associated with small oscillations around the fixed point. We show
here that the period of small oscillations is naturally around $170$
days, if the mutual inclination is near the critical value for Kozai
oscillations.

\subsubsection{The frequency of small oscillations} 
Setting the first derivative of the Hamiltonian
(\ref{eq:hamiltonian}) with respect to $\omega$ and $\G$ to zero, we
find the following expression for the location of the stable fixed
points in the limit of small eccentricity: \bea \omega_f
&=&90^\circ\ ,\ 270^\circ\\ e_f
&=&\sqrt{\frac{18-30\frac{\Ha^2}{\La^2}-{B\over A}-120k_2{C\over
      A}-3k_2{D\over A}}{60\frac{\Ha^2}{\La^2}+{3\over2}{B\over
      A}+840k_2{C\over A}+\frac{15}{2}k_2{D\over A}}}. \label{eq:efix}
\eea
We can write the second of these as 
\be %$
e_f =\sqrt{30\left[\cos^2\ic-\cos^2\imath\right]\over{60\frac{\Ha^2}{\La^2}+{3\over2}{B\over A}+840k_2{C\over A}+\frac{15}{2}k_2{D\over A}}},
\ee %$
where
\be %$
\cos^2\ic \equiv {3\over 5}-{1\over 30}{B\over A} - 4k_2 {C\over A} -{1\over 10} k_2 {D\over A}.
\ee %$

Evaluating the second derivative of the Hamiltonian at the fixed point
we obtain the expression for the frequency of small oscillation around
the fixed point:
\bea \label{eq:period} %$
\omega_0=&\omega_A&\Bigg[\left(18+90{\Ha^2\La^2\over G_f^4}\right)+2{B\over A}\frac{\La^3}{G_f^3}+k_2{C\over
    A}\left(3150\frac{\La^{11}}{G_f^{11}}-1680\frac{\La^9}{G_f^9}+90\frac{\La^7}{G_f^7}\right)+12k_2{D\over
    A}\frac{\La^5}{G_f^{5}}\Bigg]^{1/2}\nonumber\\
&\times& e_f\sin\imath_f,
\eea %$
which should be compared to equation (\ref{eq:omega_0}). As in the pure
Kozai case, the period of small oscillations $P_0\sim1/e_f$.

\begin{figure}
\epsscale{0.7} 
\plotone{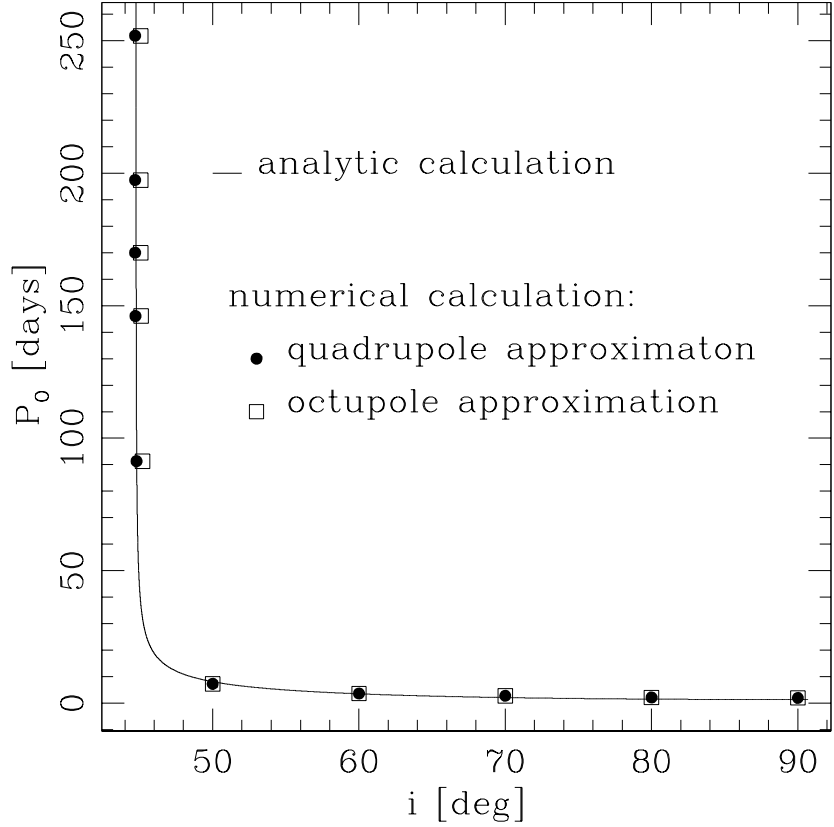}
\caption{
  The period of small oscillations vs. the initial
  inclination for a system similar to 4U 1820-30, with $k_2=0.01$. The
  critical inclination is $\ic\approx 44.7^\circ$. At large
  inclinations, well above $\ic$, the period of libration is of order
  days. Only if $\imath\approx\ic$ is the period of order $170$
  days. The solid line is the prediction of equation (\ref{eq:period}); the solid circles come from numerical integration of the equations of motion to quadrupole order, while the open squares come from integration accurate to octupole order.
\label{Fig:Pi}}
\end{figure}

Figure \ref{Fig:Pi} shows $P_0$ as a function of the initial
inclination.  As the initial inclination increases above the critical
value, the period of small oscillations decreases rapidly. Increasing
the initial inclination increases the magnitude of the Kozai torque;
in the absence of other torques, and for inclinations above the
critical inclination, increasing the magnitude of the Kozai torque is
analogous to increasing the restoring force in a harmonic oscillator,
thereby increasing the frequency of oscillation. When there are other
torques in the problem, the critical inclination will change; for
example, the presence of a tidal bulge on the secondary increases the
critical inclination.

Very near the critical inclination, the effective restoring force is
small, $\sim e_f\sin \imath_f$, so the frequency of small oscillations
is small, and the period of oscillations is large---hence the rapid
increase in $P_0$ as the inclination decreases toward the critical
inclination ($\ic\approx44^\circ$ in Figure \ref{Fig:Pi}).

\begin{figure}
\epsscale{0.7} 
\plotone{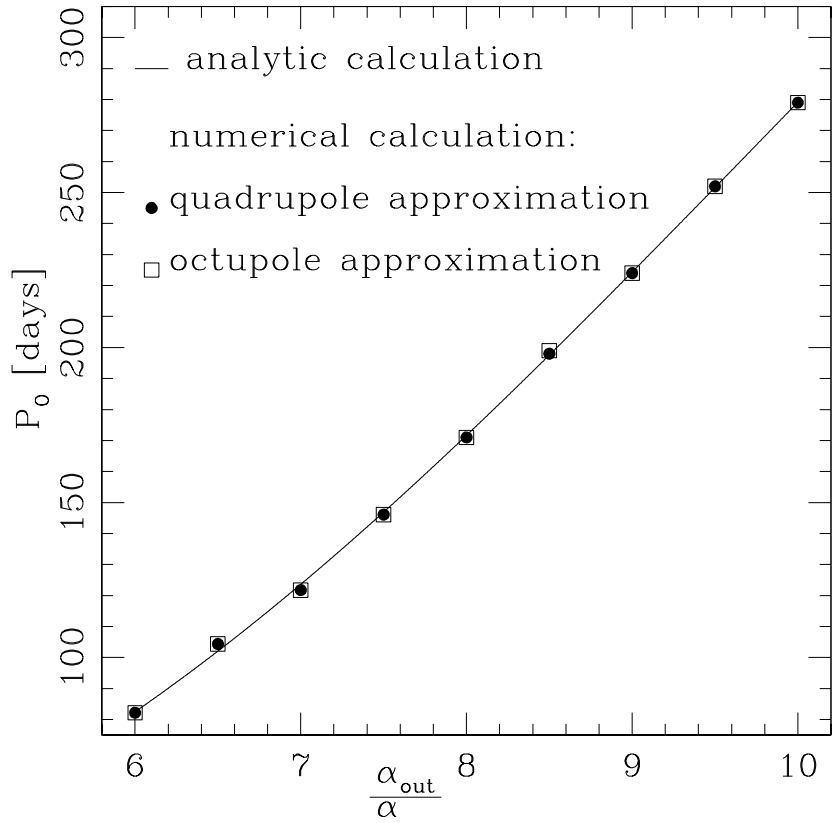}
\caption{
  The period of small oscillations vs. $a_{out}$. As the semimajor axis of the outer binary, $a_{out}$
  increases, the period of small oscillations is increasing too, which
  is expected from the $n_{out}\sim a_{out}^3$ dependance of
  $\dot\omega_{Kozai}$. The solid line is the prediction of
  equation (\ref{eq:period}), while the solid circles and open squares  are from numerical
  integrations accurate to quadrupole and octupole order respectively.
\label{Fig:Pa}}
\end{figure}

Figure \ref{Fig:Pa} shows $P_0$ as a function of $a_{out}$. As
expected from the $n_{out}\sim a_{out}^3$ dependance of $\omega_A$,
the period of eccentricity oscillations increases rather rapidly with
$a_{out}$.

\section{MASS TRANSFER, TIDAL DISSIPATION, AND CAPTURE INTO LIBRATION} \label{sec:capture}
We have shown that physically plausible values of $k_2$ lead to a
precession frequency $\dot\omega_{TB}$ that is much larger than the
observed frequency of luminosity variations in 4U 1820-30. We then
appealed to an equally large precession, of the opposite sign,
produced by the Kozai interaction, to cancel the prograde precession
caused by the tidal bulge. In order to avoid fine tuning, we argued
that the system has to be in libration, so that the observed low
frequency actually arises from libration, rather than precession of
the apsidal line of the binary orbit.

Whether the tidal bulge or GR effects produce a larger precession
rate, we argue that it is no coincidence that the magnitude of the
Kozai precession rate is equal to the sum of the other precession
rates: the system will evolve so as to capture the orbit into
resonance, in which the sum of all the precession rates is zero.

Capture into libration in the Kozai resonance is a natural consequence
of semimajor axis expansion, the latter driven by mass loss from the
white dwarf as a result of its overflowing its Roche lobe. The action
$\int \G d\omega$ is an adiabatic invariant (for detailed discussion see appendix B), since the semimajor axis
of the binary orbit is expanding on the accretion time scale $m_2/\dot
m_2\approx 10^7\yr$, much greater than either the orbital or
precession time scale. In contrast to mass transfer, tidal dissipation
tends to shrink the semi-major axis; if this effect dominates,
trapping into the Kozai resonance is not possible.

How does expansion of the inner orbit lead to capture into libration?
As $a$ increases, the mutual torque between the two orbits will
increase as well---the inner orbit is expanding, effectively moving
closer to the outer orbit. This increasing torque corresponds to a
deepening of the Kozai potential, and an expansion in the size of the
separatrix of the Kozai resonance. Orbits other than the separatrix
have a fixed action, while the action of the separatrix is
increasing. If the increase in the action of the separatrix grows to
exceed the action of an initially circulating orbit, that circulating
orbit will be captured into resonance, and begin to librate. As $a$
continues to expand, the captured orbit will move closer and closer to
the fixed point of the resonance, librating with the frequency of
small oscillations.

More quantitatively, mass transfer tends to increase $a$ \citep{1982ApJ...254..616R}:
\be %$
\dot a_{MT}={1\over2/3-1/q}{3\times2^{23/6}\over5c^5}
\Big[{K\over0.4242}\Big]^{3/2}
{m_1(G(m_1+m_2))^{3/2}\over a^{9/2}},
\ee  %$
where $q=m_1/m_2$ and $K=k\theta_{\gamma'}/(\mu m_p)$; $k$ is the
Boltzmann constant, $\mu$ is the mean molecular weight, $m_p$ is the
mass of the proton and $\theta_{\gamma'}$ is the polytropic
temperature. The parameter $K$ is given by the following mass-radius
relation: \be K=N_nGm_2^{1-(1/n)}R^{(3/n)-1}, \ee where $N_n$ is a
tabulated numerical coefficient (for $n=1.5$ it is 0.4242;
\citet{1939isss.book.....C}).

Tidal dissipation in the white dwarf will tend to reduce the semimajor
axis of the binary. In the limit of small eccentricity,
\be \label{eq:tidal dissipation}%$
\left({da\over dt}\right)_{TD}\approx-{38\over3}na{k_2\over Q}{m_1\over m_2}
\left({R_2\over a}\right)^5 e^2 .
\ee %$

We argue that the orbit must be expanding.  $(e/ \dot e )_{TD}$ is $~100$ times shorter  than $(a/ \dot a )_{TD}$, so unless something excites $e$ (such as third body or thermal tides) we are unlikely to catch the system in a phase where periastron, $r_p=a(1-e)$, is increasing while $a$ is decreasing.

\section{NUMERICAL RESULTS}\label{sec:numerics}
\subsection{Numerical model using the quadrupole approximation}

Our numerical model treats the gravitational effects of the
third body in the quadrupole approximation. We average over the orbital
periods of both the inner binary and the outer companion. We demonstrate in section \ref{sec:octupole} and in figures \ref{Fig:Pi} and \ref{Fig:Pa} that treating the effects of the third body in octupole approximation does not qualitatively change our findings. We include
the following dynamical effects:
\begin{itemize}
\item periastron advance due to general relativity;
\item periastron advance arising from quadrupole distortions of the
  helium white dwarf due to both tides and rotation;
\item orbital decay due to tidal dissipation in the white dwarf;
\item loss of binary orbital angular momentum due to gravitational radiation;
\item conservative mass transfer from the helium white dwarf to the neutron
  star primary driven by the emission of gravitational radiation.
\end{itemize}
\noindent Note that the Kozai mechanism described in the previous
section is included in the three body gravitational dynamics. The
equations used in our model are listed in the appendix.

\subsection{Results}

We use as fiducial parameters $m_1=1.4M_\odot$, $m_2=0.067M_\odot$,
and $m_3=0.55M_\odot$. The semimajor axis of the inner binary is
$a=1.32\times10^{10}\cm$, chosen to match the observed orbital period
of $685\s$. The radius of the Helium white dwarf is
$R_2=2.2\times10^9\cm$, while the fiducial Love number is $k_2=0.01$.

\begin{figure}
\epsscale{0.6} 
\plotone{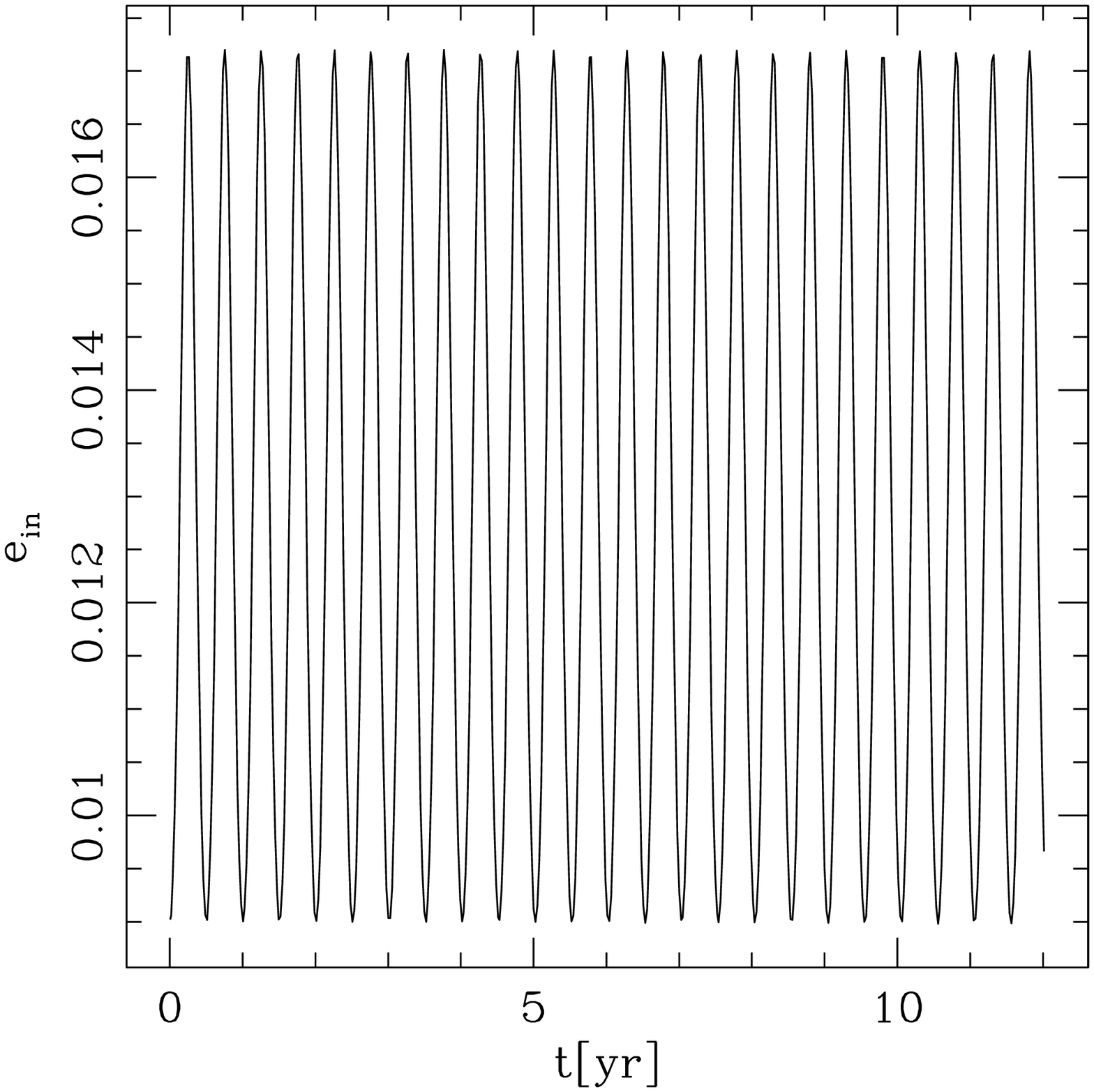}
\plotone{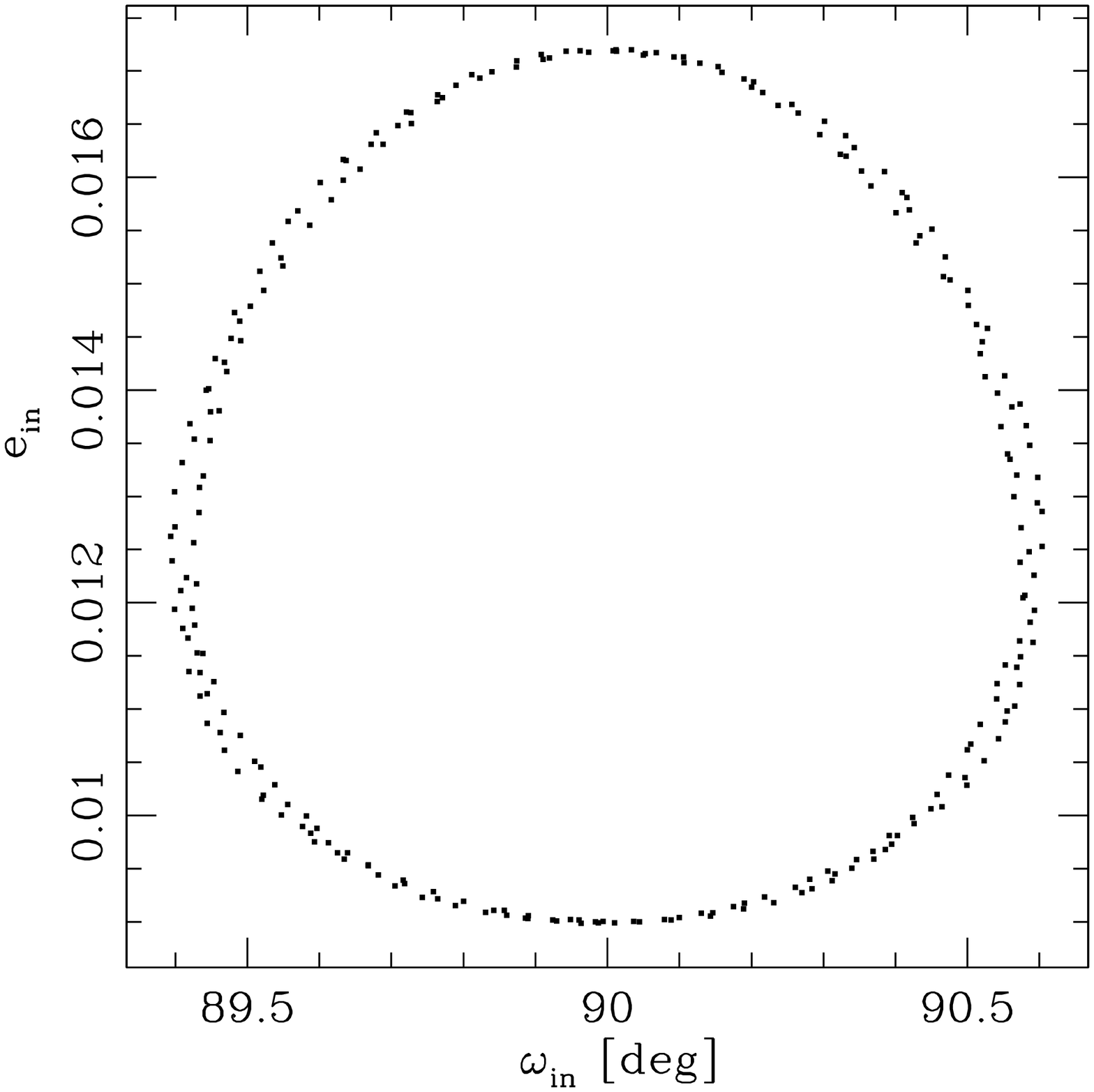}
\caption{
  The eccentricity as a function of time (upper
  panel) and the phase space ($e$ versus $\omega$) for our fiducial
  model. The period of the eccentricity oscillations is $171$ days,
  and the amplitude of the eccentricity oscillation is sufficient to
  produce the observed factor of $2-3$ variation in luminosity.
\label{Fig:ecc}}
\end{figure}

To reproduce the $171$ day eccentricity oscillations (Figure
\ref{Fig:ecc}), we use the following initial parameters:
$a_{out}=8.0a$ (yielding $P_{out}=0.15$ day). We start with
$e_0=0.009$, $\omega_0=90^\circ$, $\imath_{init}=44.715^\circ$, and
$e_{out,0}=10^{-4}$.

\begin{figure}
\epsscale{0.7} 
\plotone{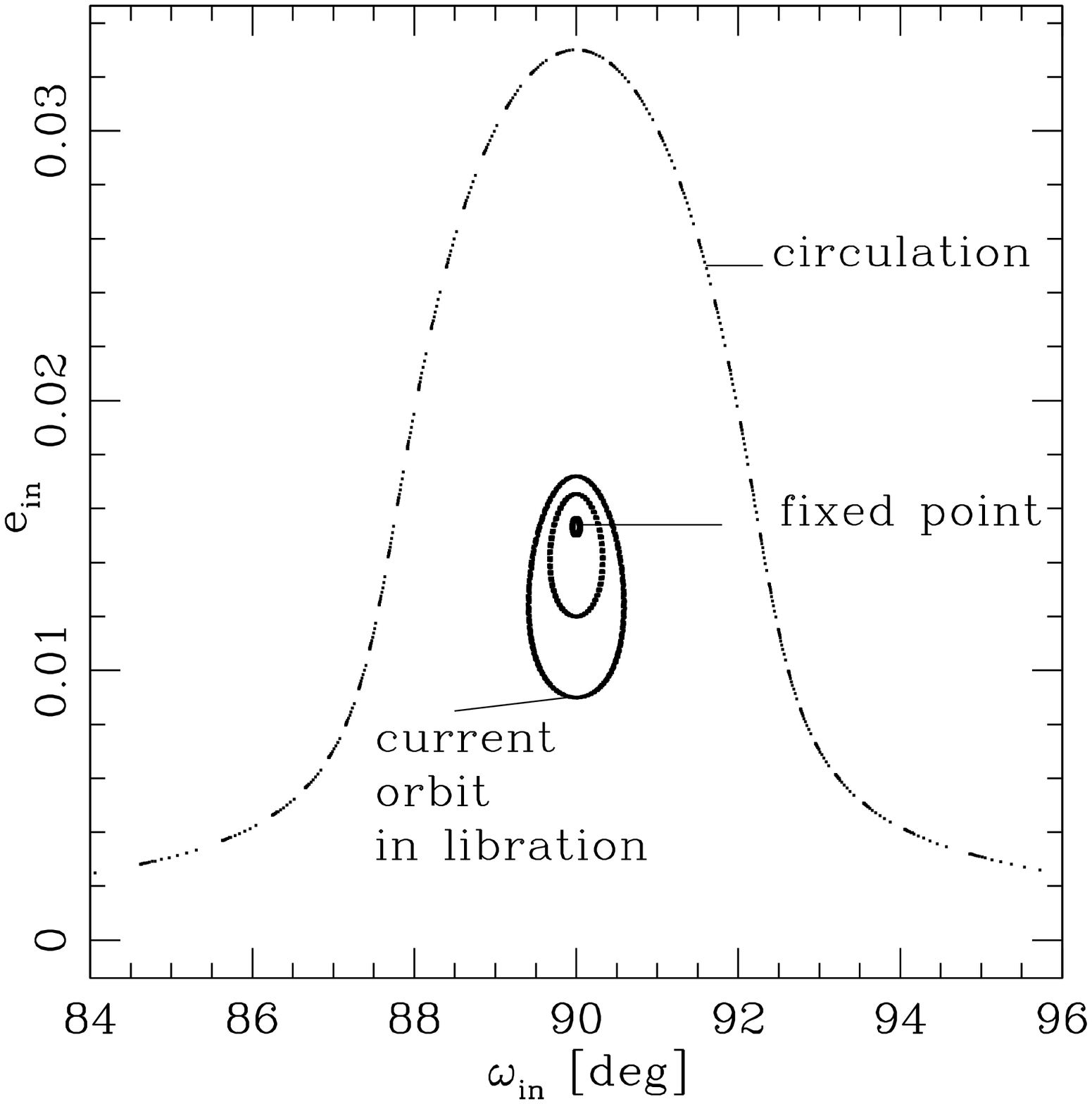}
\caption{
Phase portrait for four different initial
eccentricities at initial inclination $\imath=44.715^o$ and initial
$\omega=90^\circ$. Our fiducial orbit, with
a libration period of $171$ days and the amplitude of the eccentricity oscillations sufficient to produce observed variations in the luminosity, is labeled as the ``current orbit in libration''; it is the same orbit presented on
Figure \ref{Fig:ecc}. The unlabelled librating orbit is very near the fixed point; it has a period of the eccentricity oscillations that is close to but  shorter than the observed period, while the variations induced in the luminosity are too small compared to the observations. The
circulating orbit produces luminosity variations that are too large, as well as having an oscillation period that is too long.
\label{Fig:ps_all}}
\end{figure}

Figure \ref{Fig:ecc} shows the eccentricity oscillations of the inner
binary, with a period of $~171$ days, over a decade. The amplitude of
the eccentricity oscillations is of order of $7\times10^{-3}$, which
is sufficient to enhance mass transfer enough to produce the observed
luminosity oscillations of a factor of $\gtrsim2$ \citep{2007MNRAS.377.1006Z};
see their Figure 3.  
The amplitude of the eccentricity oscillations depends on the initial
eccentricity, as illustrated in Figure \ref{Fig:ps_all}; a lower
initial eccentricity produces eccentricity oscillations with higher
amplitude. If the system circulates, the amplitude of the eccentricity
oscillations is larger still.

Having the system trapped in libration about the fixed point explains
both the origin of the $171$ day period luminosity variations, as well
as the small amplitude of the eccentricity oscillations; the
observations require that magnitude of the eccentricity oscillations
be small so as to avoid overly large luminosity variations---a point
we return to below.

%\clearpage
%\clearpage

%\clearpage
%\clearpage 

\subsection{Resonant trapping and detrapping of 4U 1820-30}

\noindent 

The mass transfer rate is determined by the inner binary mass and
semimajor axis. These parameters are reasonably well constrained from
observations \citep{1987ApJ...312L..17S, 1997ApJ...482L..69A,
  1987ApJ...322..842R}. The amount of tidal dissipation is
parameterized by the tidal dissipation factor $Q$, which for white
dwarfs is not well constrained at all. If we know the value of
period derivative, $\dot P$, we can constrain $Q$ (or more precisely,
$(e/0.009)^2Q/k_2$, see equation \ref{eq:tidal dissipation}) for the white dwarf
in the system.

\begin{figure}
\epsscale{0.49} 
\plotone{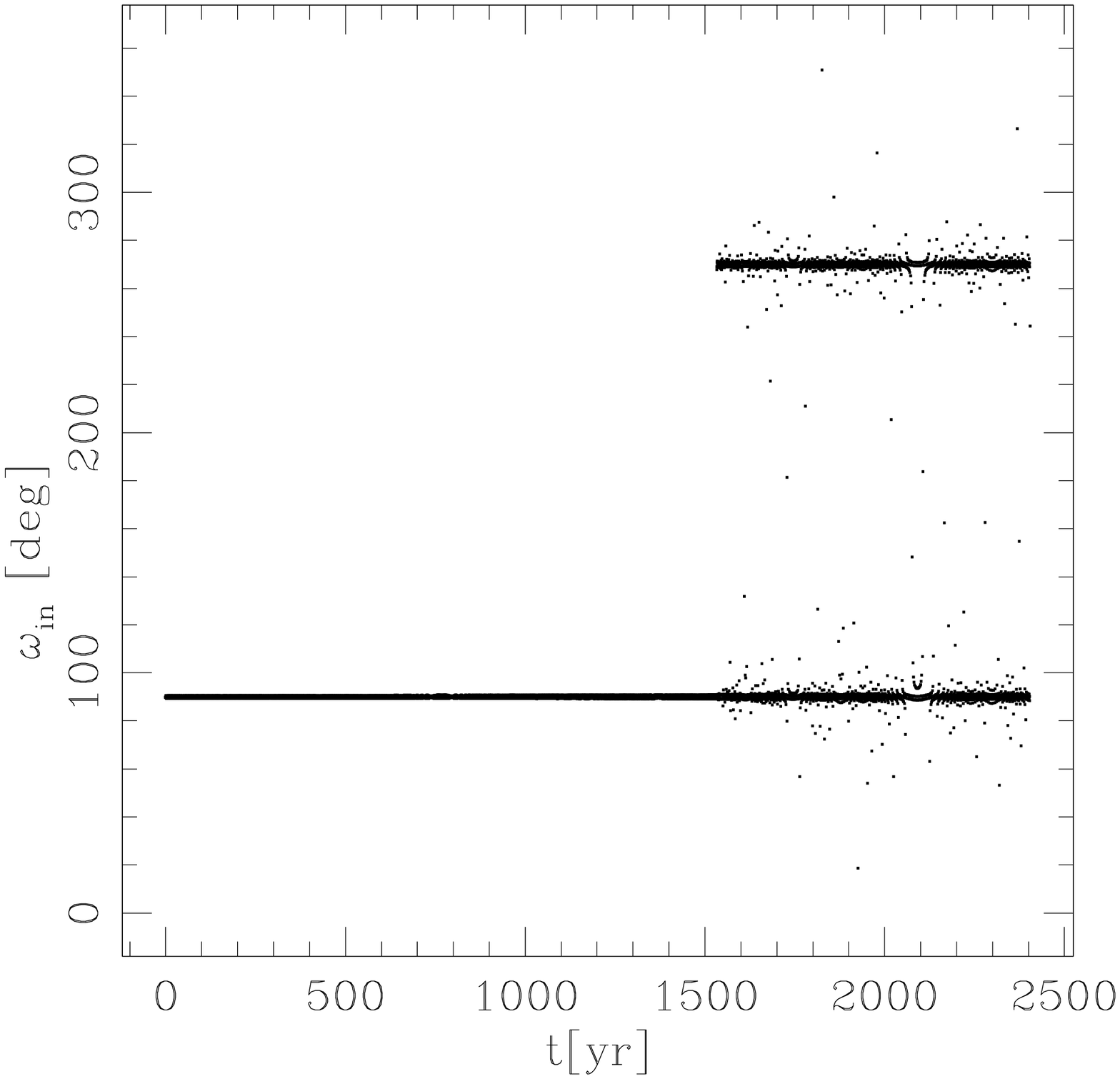}
\plotone{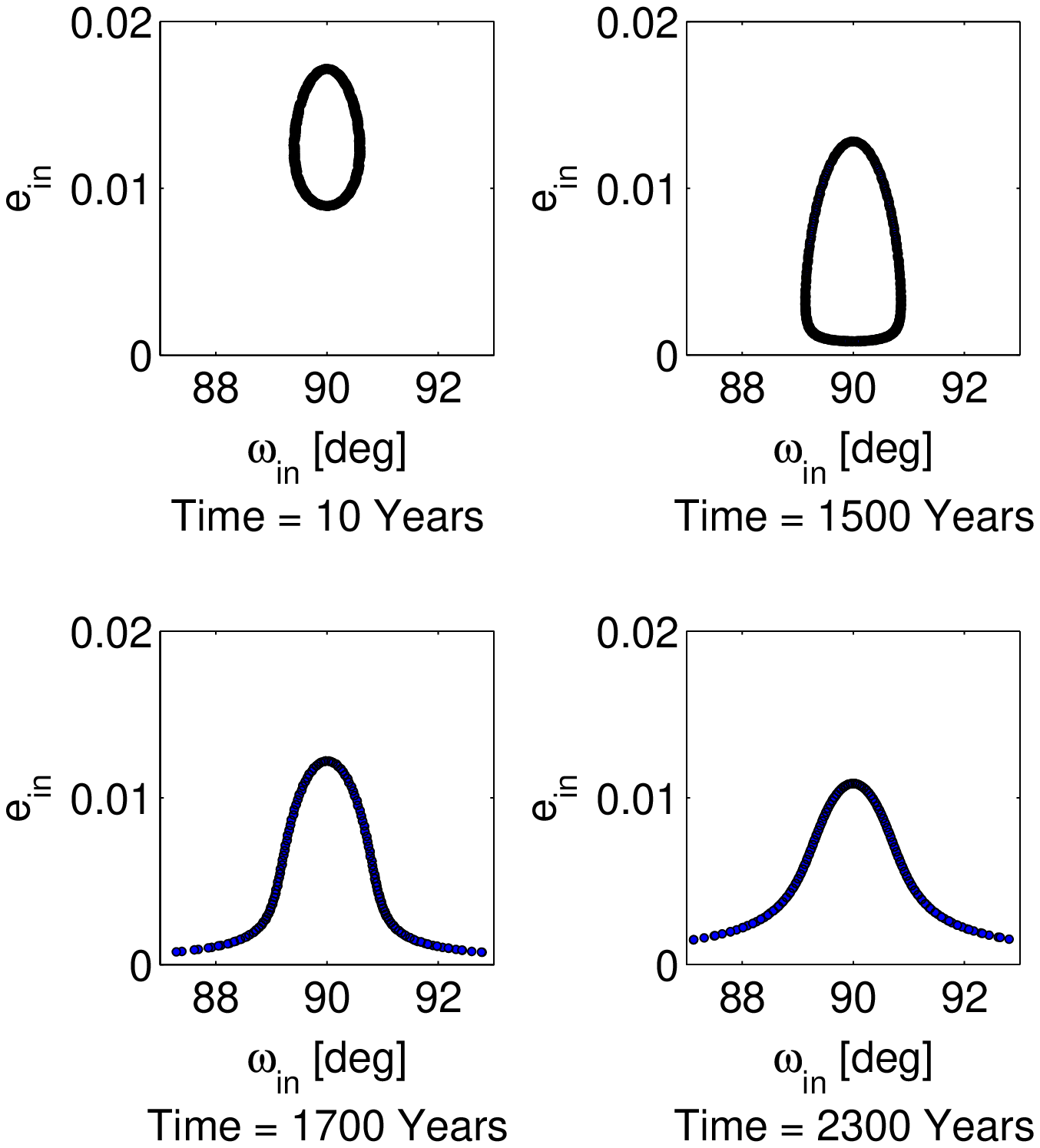}
\plotone{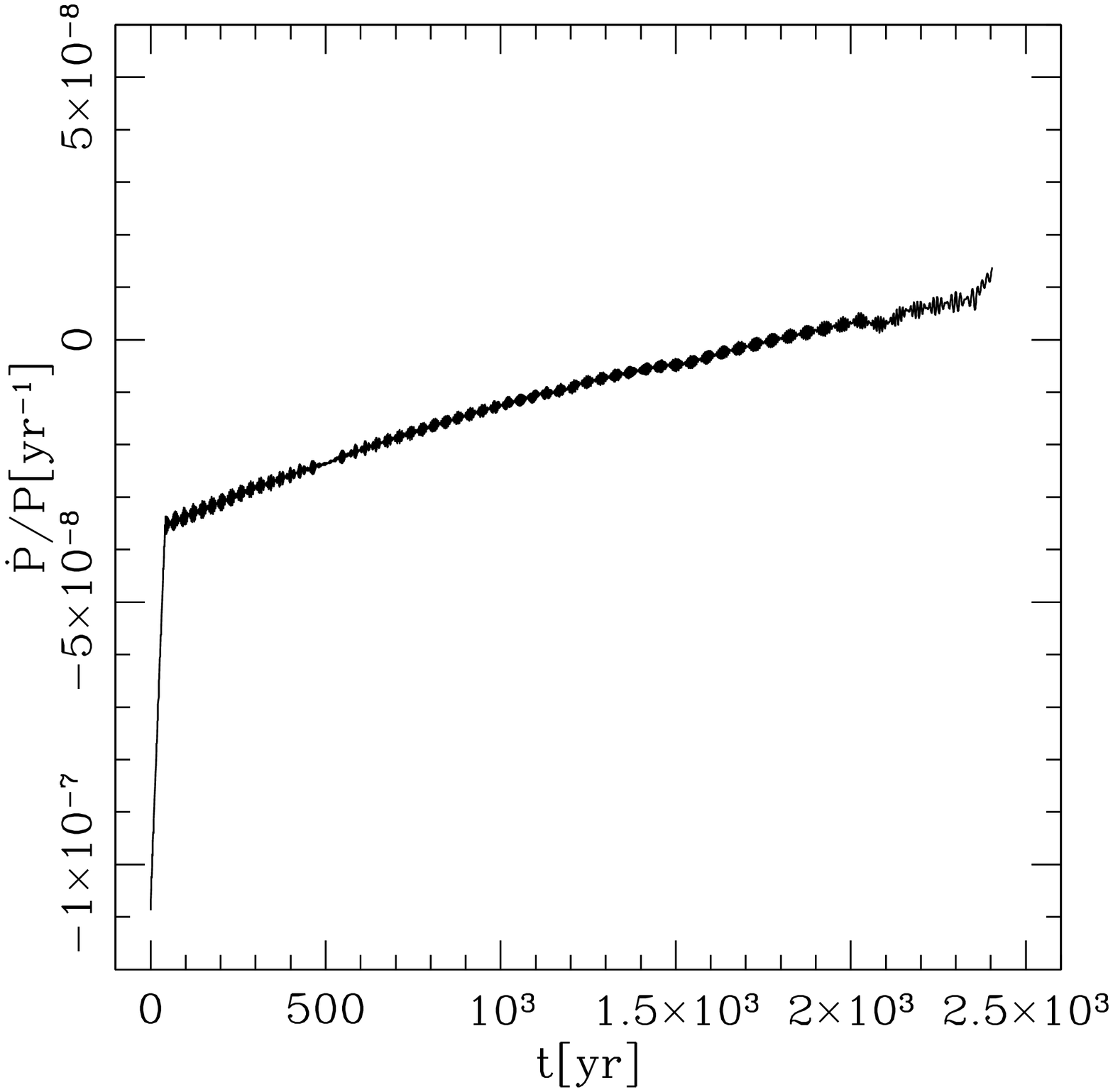}
\caption{
a) $\omega$ vs $t$. We start the evolution of the system by placing
the system in libration. Here we use $Q=2.5\times10^{7}$, which is the value
required to reproduce the observed negative period derivative. The
action of the separatrix is decreasing, and the system is ejected from
the resonance after about $1500\yr$. b) $e$ vs $\omega$, phase space
evolution plot, showing that the orbit evolves from libration to
rotation, with the transition occurring between the $1500$ and $1700\yr$ snapshots. c) $\dot{P}/P$ vs $t$. The period derivative remains negative only for about $1700\yr$, which is $10^{-3}$ of the lifetime of the system; the eccentricity damps sufficiently that the $\dot m$ term takes over.
\label{Fig:kick_out}}
\end{figure}

We argued at the end of \S \ref{sec:capture} that the intrinsic $\dot
P$ must be positive, since a shrinking binary orbit and a decaying eccentricity quickly lead to mass transfer driving expansion of the binary orbit. There is a second argument
against an intrinsic negative $\dot P$: if the orbit of the inner
binary is shrinking, an initially librating orbit will quickly become
circulating, and the period of luminosity variations will change
dramatically. If we tune $Q$ to the value that reproduces the observed
negative period derivative ($Q=2.5\times10^7$, assuming $k_2=0.01$) and let the
system evolve, the system is driven out of libration after about
$~1500\yr$, as shown in Figure \ref{Fig:kick_out}. As the figure shows, the eccentricity of the inner binary decreases significantly due to tidal dissipation, which in turn reduces the strength of tidal dissipation. With tidal dissipation weakened, mass transfer will dominate the evolution of the semimajor axis and, as expected from the standard evolutionary scenario, the semimajor axis starts to expand (not shown in the figure). As long as there is some small eccentricity in the inner binary there is some tidal dissipation present that tends to slow down the expansion rate of the semimajor axis.  

The reason for the detrapping is rather subtle. First, we note that the decrease in $e$ is \underline{not} due to direct tidal damping; Equation (\ref{eq:e_dot_TD}) predicts $(e/\dot e)_{TD} \sim 10^5 \yr$, while $e$ changes by factor of $2$ in $~2000 \yr$. To verify this, we have set $\dot e_{TD}=0$, and verified that integration yields the same result. The reason for such a short time scale for decrease in $e$ lies in the fact that the spins do not remain tidally locked throughout the evolution of the system and the evolution of the eccentricity is rather strongly influenced by the their lack of pseudo-synchronism. Detailed discussion and figures are given in appendix B.

%\clearpage
%\clearpage  

On the other hand, if the observed negative period is not an intrinsic
property of the system, in other words, if the effect of mass transfer
wins over the effect of tidal dissipation, the action of the
separatrix increases with time, and trapping will occur.

\begin{figure}
\epsscale{0.55}
\plotone{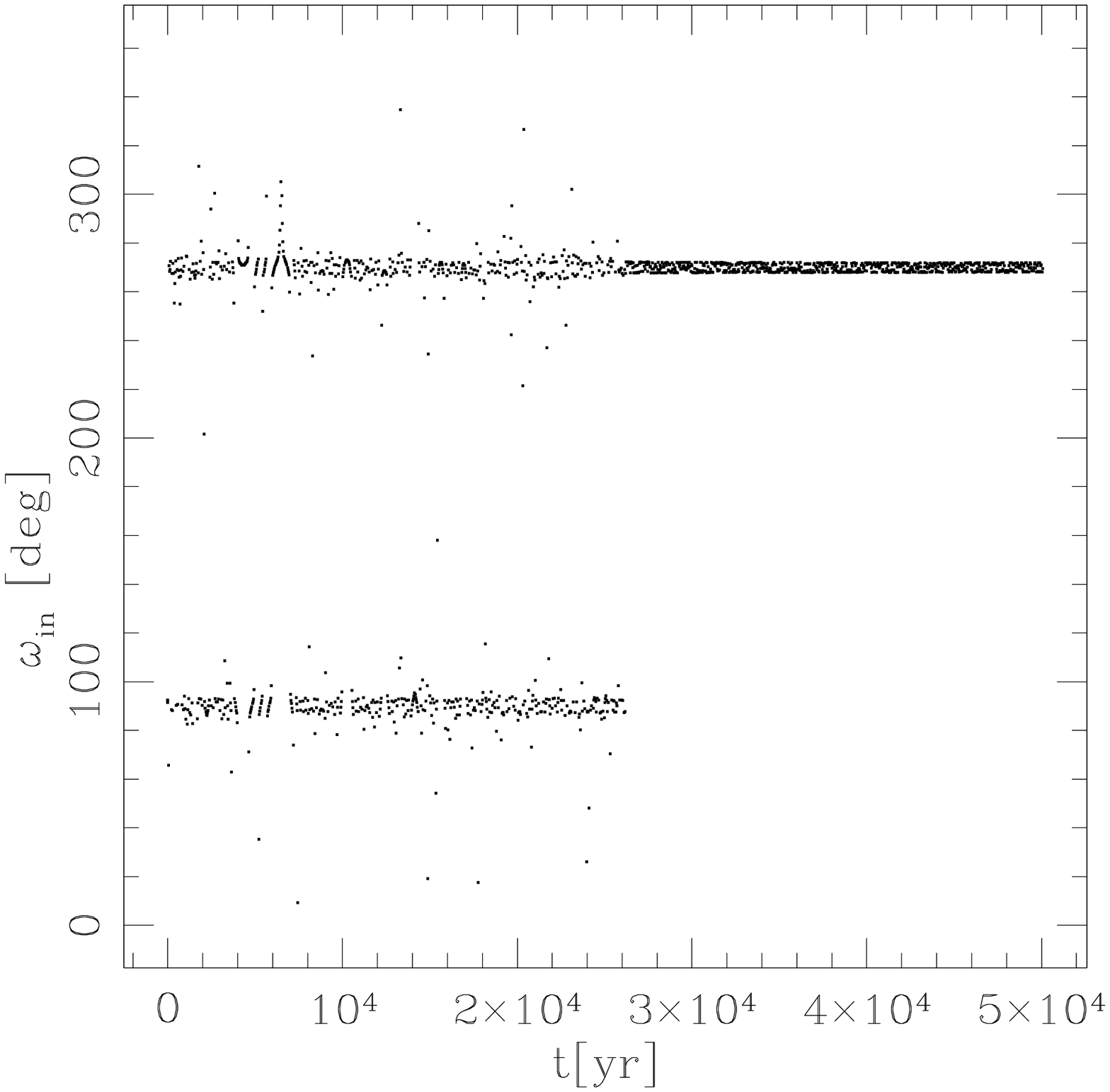}
\plotone{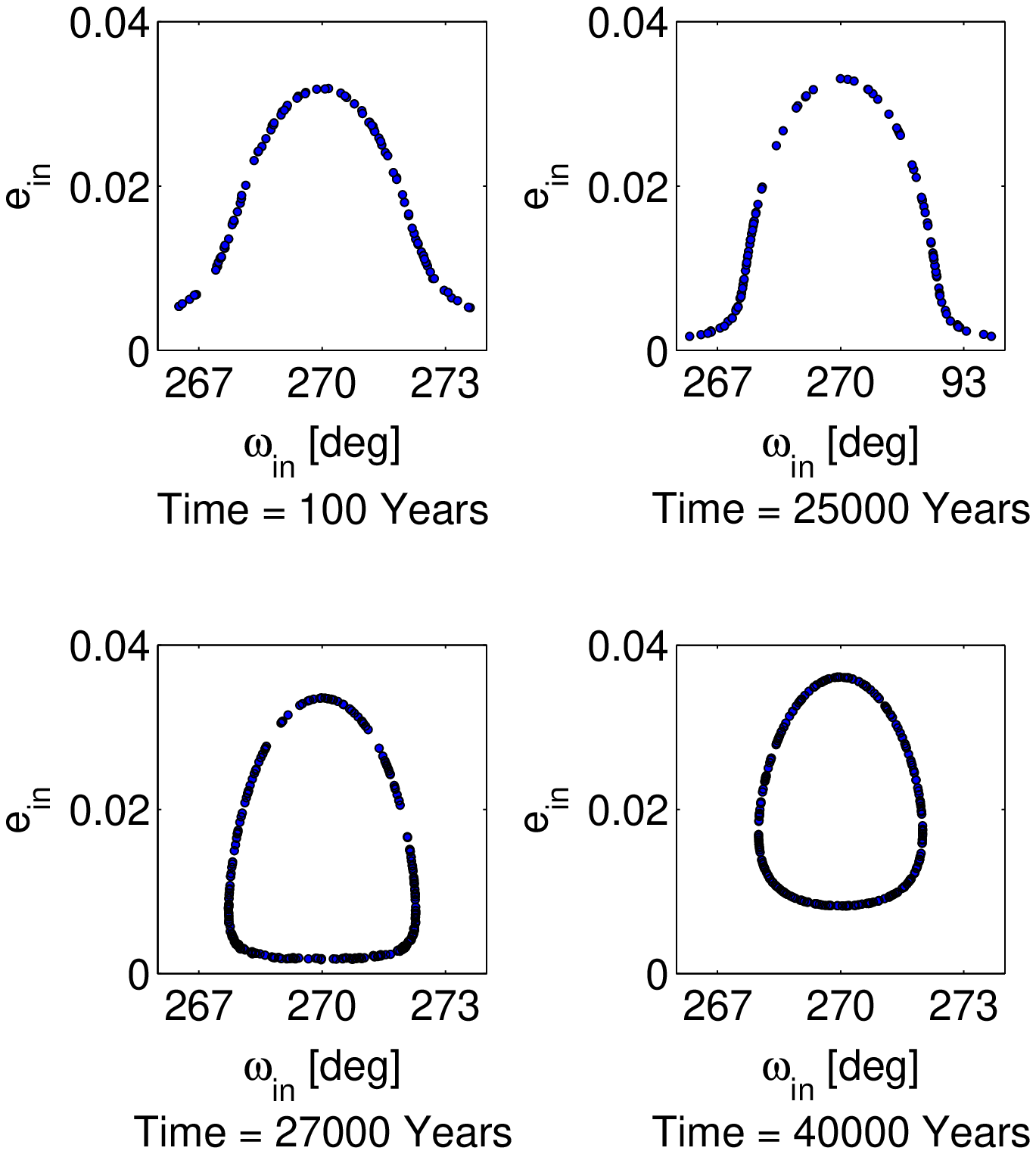}
\caption{
  a) $\omega$ vs $t$. The system is placed in
  circulation; after about $27000\yr$ it gets trapped in libration.
  Here we have used $Q=5\times10^7$, while the initial eccentricity is $e_0=0.032$; all
  other parameters are the same as used in Figure
  \ref{Fig:ecc}. b) $e$ vs $\omega$ for the same integration.
\label{Fig:trap_omega}}
\end{figure}

Figure \ref{Fig:trap_omega} shows a system initially put on a
circulating orbit. As the integration proceeds, the separatrix
expands, eventually capturing the orbit, which then librates for the
duration of the integration.

%\clearpage
%\clearpage 

\subsection{Numerical model using octupole approximation}\label{sec:octupole}

\begin{figure}
\epsscale{0.6} 
\plotone{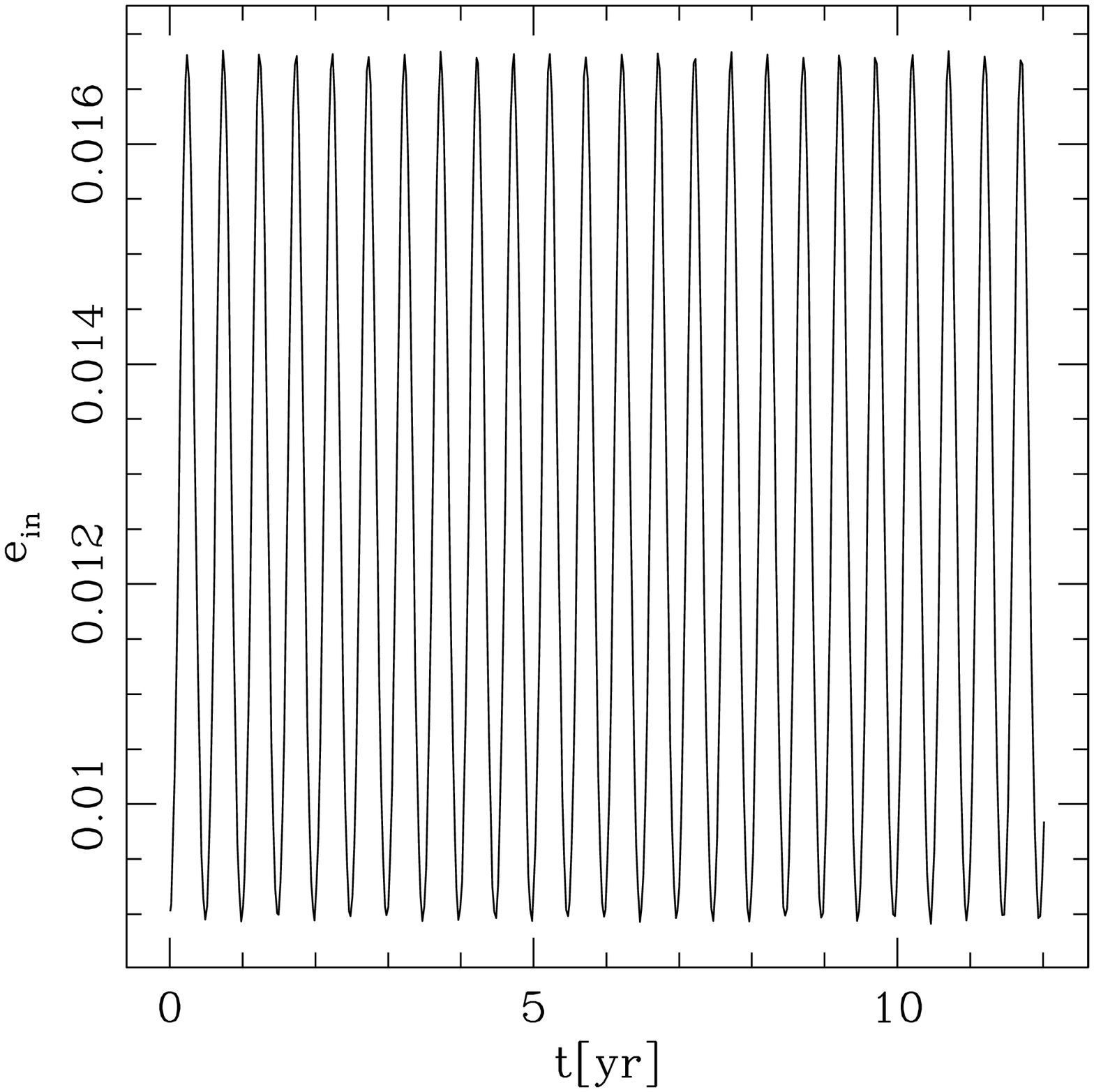}
\plotone{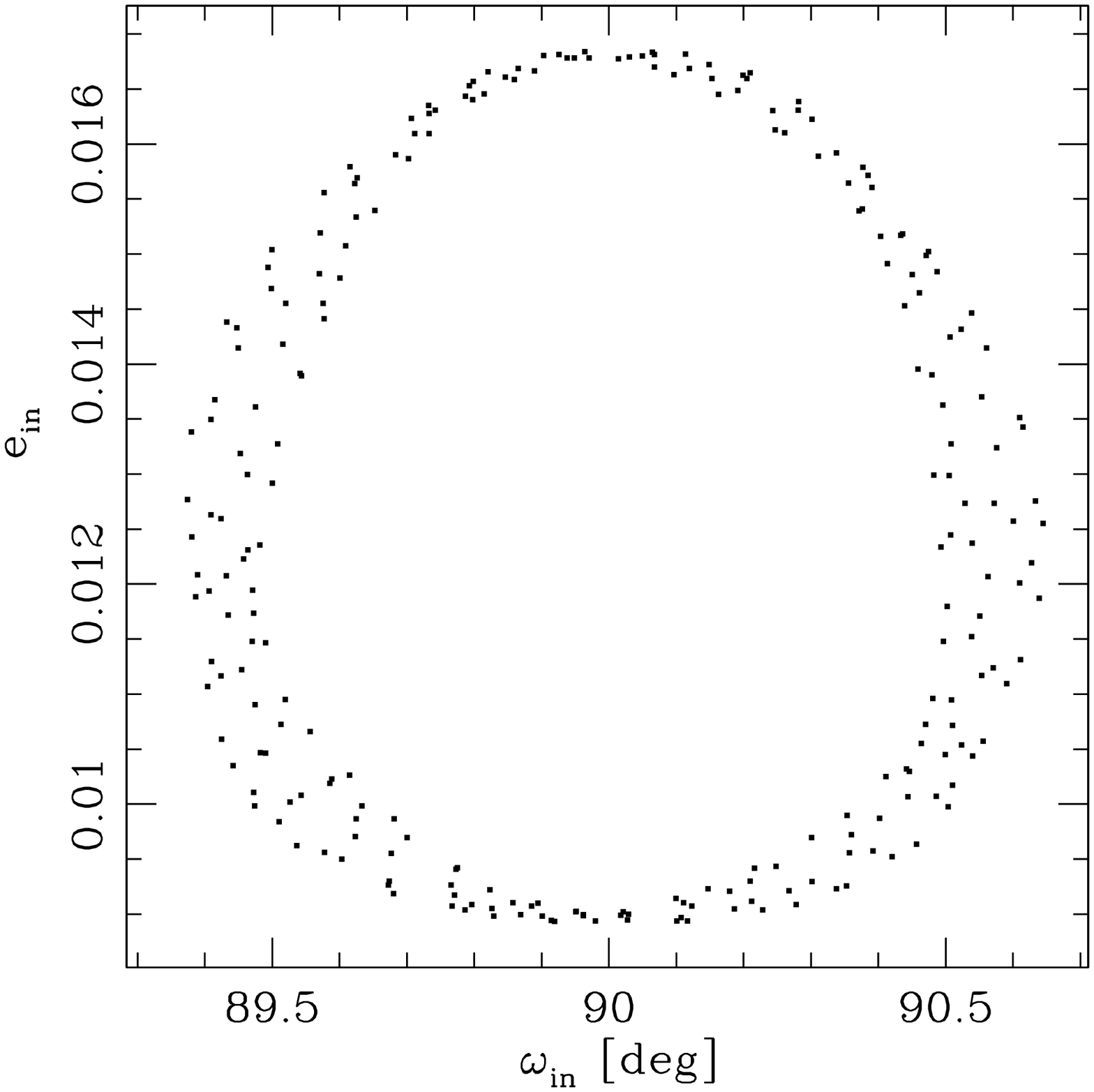}
\caption{
  The eccentricity as a function of time (upper
  panel) and the phase space ($e$ versus $\omega$) in the octupole approximation. To produce  the  $171$ days period of the eccentricity oscillations,
  and the amplitude of the eccentricity oscillation that is sufficient to
  produce the observed factor of $2-3$ variation in luminosity, slightly higher inclination is required ($\imath = 45.1^o$).   
\label{Fig:ecc_oct}}
\end{figure}

In this subsection we treat gravitational effects of the third body in the octupole approximation. As in the case of the quadrupole approximation, we derive our equations of motion from the double averaged Hamiltonian and we include all of the previously listed  dynamical effects.  As Figure \ref{Fig:ecc_oct} demonstrates, the octupole approximation does not change qualitatively our previous findings. All parameters, except the initial inclination,  used in the octupole approximation are listed in table 1. In order to produce  the  $171$ days period of the eccentricity oscillations,
  and the amplitude of the eccentricity oscillation that produces the observed factor of $2-3$ variation in luminosity, a slightly higher inclination is required ($\imath = 45.1^o$).

%\clearpage
%\clearpag

\section{ON THE VALUE OF $Q$ AND THE ORIGIN OF THE SMALL (OR NEGATIVE) $\dot P$} \label{sec:constraints}
The standard theory of Roche lobe overflow predicts $\dot
P/P\ge+8.8\times10^{-8}\yr^{-1}$. The measured $\dot
P/P=(-3.47\pm1.48)\times10^{-8}\yr^{-1}$ is eight standard deviations
away from this value. We have argued in previous section that $\dot
P/P$ should be positive, but even if it is two or three standard
deviation from the measured value, it is still five below the
predicted value. The origin of this discrepancy has been a puzzle
since it was discovered.

The suggestion that the binary has a finite eccentricity immediately
suggests a reason for the low value of $\dot P$: tidal dissipation in
the white dwarf will tend to reduce the semimajor axis of the orbit,
contributing a substantial negative term to $\dot P$.

The tidal dissipation could in fact dominate the orbital evolution,
overcoming the effects of mass transfer as seen in Figure \ref{Fig:kick_out}. We do not argue for this point of view, however, because it would be unlikely that the system could be observed in a stage of the evolution that last only $10^{-3}$ of its lifetime. In addition, we believe that the system is trapped in libration. 

The observed $\dot P/P$ consists of at least three parts:
\be \label{eq: pdot}%$
\left({\dot P\over P}\right)_{obs} =
\left({\dot P\over P}\right)_{Roche}+
\left({\dot P\over P }\right)_{accel}+
\left({\dot P\over P}\right)_{TD}.
\ee %$
The values of the observed and Roche terms were given above, and, as noted there, they are not consistent with each other. The second term on the right hand side of equation (\ref{eq: pdot}) represents the acceleration experienced by 1820-30 in the gravitational field of its host globular cluster, while the third term on the right represents the effects of tidal dissipation in the white dwarf secondary. 

A natural explanation for the observed negative $\dot P$ might be provided by a combination of the last two effects, but still allow for the system to be trapped in resonance. First, tidal dissipation reduces the intrinsic $\dot P/P$ substantially from
that expected due to Roche lobe overflow alone, but leaves $\dot
P/P>0$. We then we appeal to the argument of \citet{1993MNRAS.260..686V},
that the $(\dot P/P)_{accel}$ term produces an {\em apparent} negative total $\dot P$. Indeed, given the most recent published estimate of $a_{max}/c=7.9\times10^{-8}yr^{-1}$ from \citet{1993MNRAS.260..686V}, it is plausible that we would observe a negative period derivative, while the intrinsic (or physical) period derivative is in fact postive.

However, recent estimates for the cluster acceleration from millisecond pulsar timing suggest a maximum of $a_{max}/c=1.3\times10^{-9}yr^{-1}$ (Lynch and Ransom, private communication), an order of magnitude smaller than the estimate from \citet{1993MNRAS.260..686V}; if the smaller value holds up, the observed negative period derivative is difficult to understand in the context of current models.

\begin{figure}
\epsscale{0.55} 
\plotone{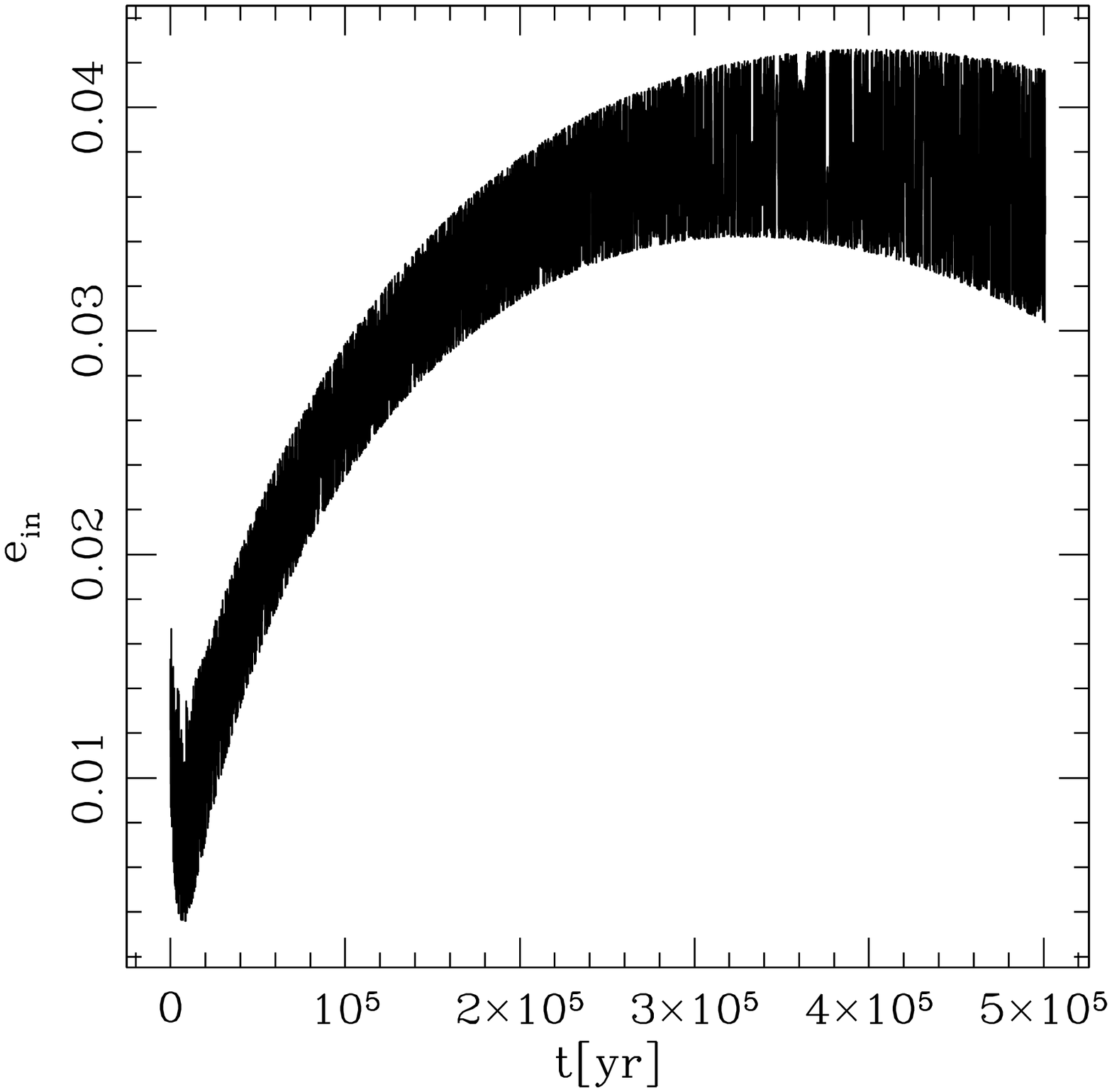}
\plotone{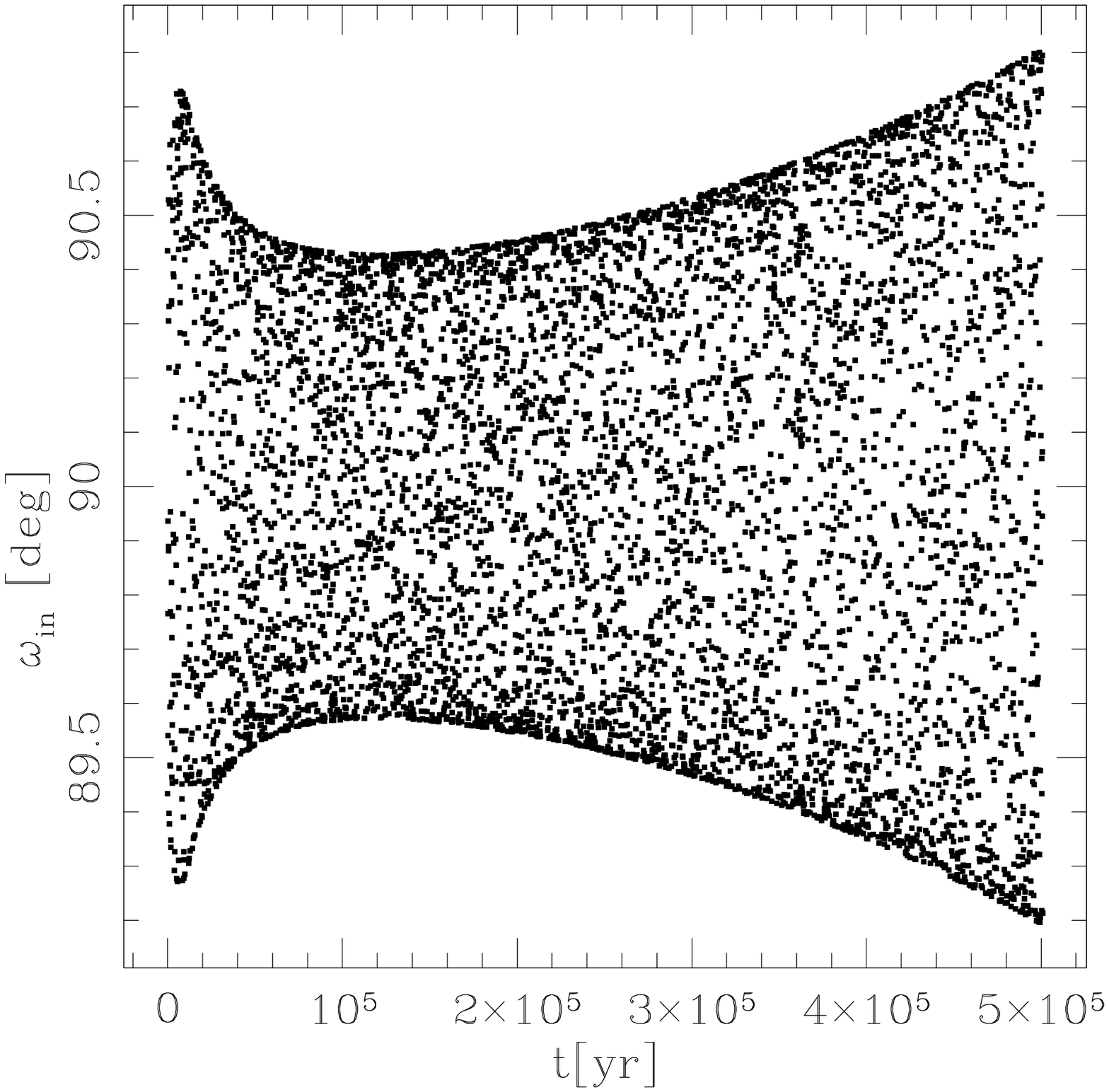}
\caption{
  The eccentricity as a function of time (upper
  panel) and the argument of periastron as a function of time ( $\omega$ versus $t$, lower panel) in the quadrupole approximation using $(e/0.009)^2Q/k_2=4.5\times10^9$. The system remains trapped in the resonance for more than $10^5\yr$ which is a considerable fraction of the system lifetime. The eccentricity stays under the limit of $0.07$, a constraint imposed by the absence of $L_2$ mass loss.    
 \label{Fig:detrap}}
\end{figure}

\begin{figure}
\epsscale{0.55} 
\plotone{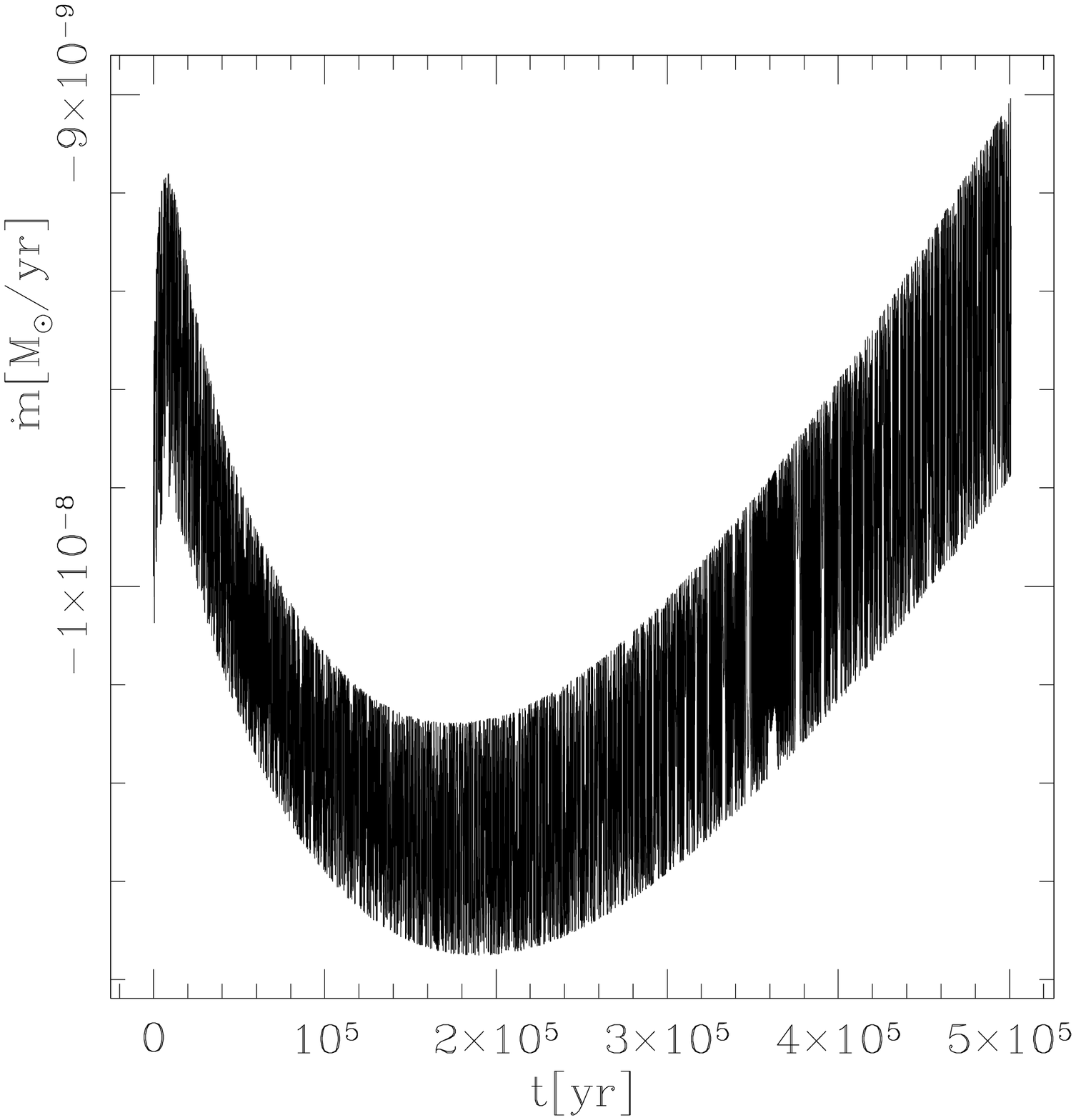}
\plotone{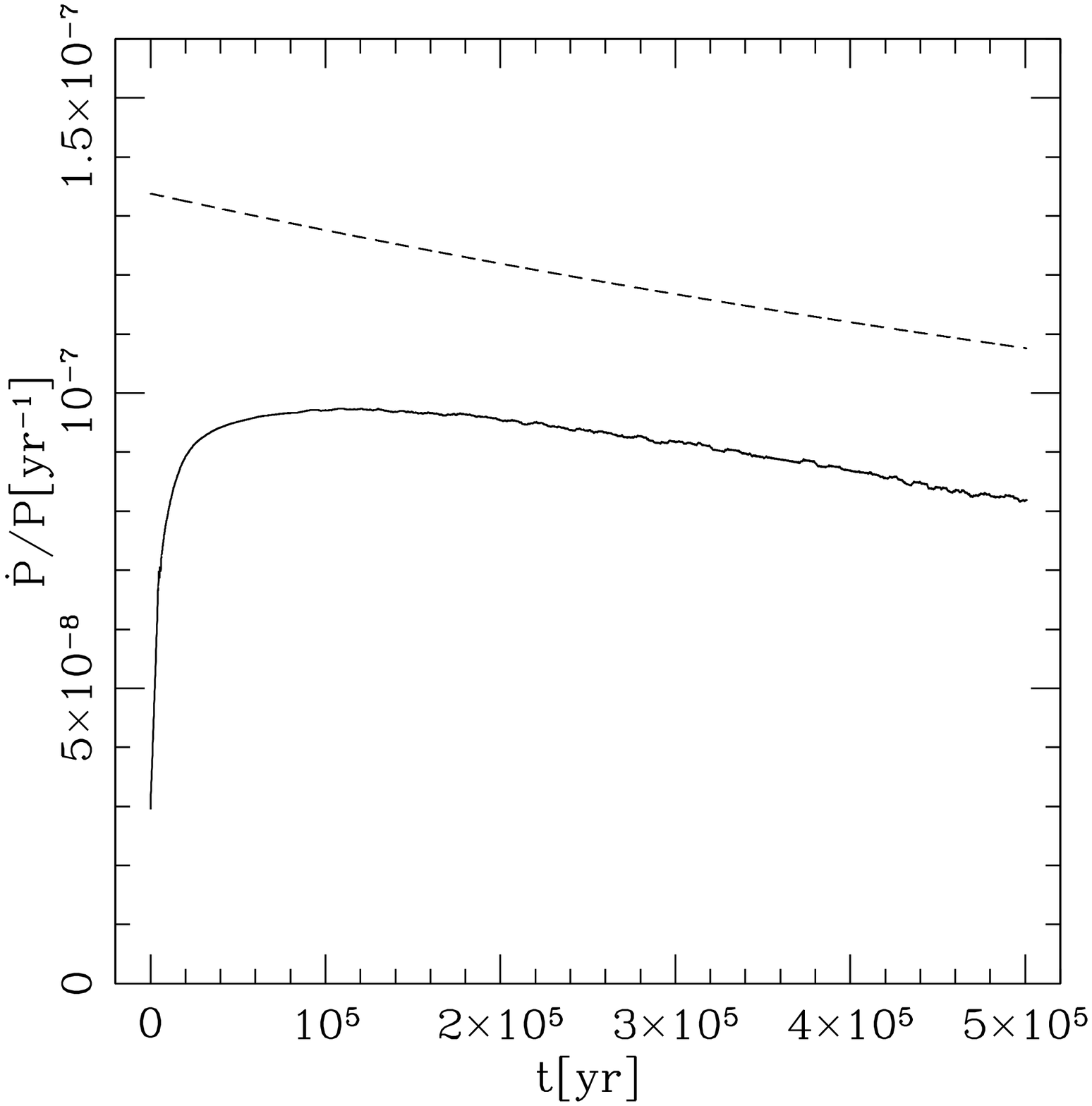}
\caption{
  The mass transfer rate as a function of time (upper panel) and $\dot{P}/P$ (lower panel, solid line) as a function of time in the quadrupole approximation using $(e/0.009)^2Q/k_2=4.5\times10^9$. The system remains trapped in the resonance for more than $10^5\yr$, which is a reasonable fraction of the system lifetime. The mass transfer rate is within $~10\%$ of its nominal value $\dot{m}_2 \thickapprox10^{-8}\mpy$. $\dot{P}/P$ is lower than that due to Roche lobe overflow alone (dashed line), but still $>0$.
\label{Fig:mt_rate}}
\end{figure}

Given that the measured negative period dervative is significant only at the two-sigma level, and that there is no clear physical explanation for such an orbital decay, it is worth considering the possibility that the observed value is in error. If we ignore the observed negative period derivative, and simply assume that the intrinsic $\dot P$ is postive, we find a lower limit on $Q$  given by $(e/0.009)^2Q/k_2>3.15\times10^9$.  We can get a firmer lower limit on $Q$ by  requiring the system to remain trapped in a resonance for a considerable fraction of its lifetime. Given that $m_2=0.067M_\odot$ and $\dot m_2\approx 10^{-8}\mpy$, the lifetime during which this system can sustain its high X-ray luminosity is estimated to be $~7$ million years, so a reasonable fraction of its lifetime to remain trapped in a resonance is at least $10^5\yr$. According to our model for $(e/0.009)^2Q/k_2> 4.0\times10^9$ the system remains trapped in the resonance for more than $10^5\yr$ (see Figure \ref{Fig:detrap}), and as Figure \ref{Fig:mt_rate} demonstrates, the mass transfer rate remains within roughly $10\%$ of its nominal value $\dot{m}_2\thickapprox10^{-8}\mpy$. In this case, the eccentricity of the inner binary will never damp down to a fixed point because it is indirectly driven up by semimajor axis expansion due to mass loss on a timescale of order $~10^4\yr$, which is at least an order of magnitude shorter then the timescale for eccentricity damping due to tidal dissipation. As expected from the evolutionary scenario the intrinsic period derivative is positive, but because of the effect of tidal dissipation it is smaller than that  due to Roche lobe overflow alone (see Figure \ref{Fig:mt_rate}); the nominal value for the period derivative due to Roche lobe overflow alone for our system parameters is $\dot{P}/P\thickapprox 1.3\times10^{-7}\yr^{-1}$ \citep{1987ApJ...322..842R}.
 
Finally, we note that if the inner binary is in fact expanding, the eccentricity will tend to increase as well. If the eccentricity is large enough, then Roche lobe overflow will occur through both the inner and outer Lagrange points, in contradiction with the low observed x-ray absorption. Figure \ref{Fig:detrap} shows that the eccentricity, while increasing with time, remains smaller than $0.07$, consistent with the lack of mass loss through the outer ($L_2$) Lagrange point \citep{2005MNRAS.358..544R}.

%\clearpage
%\clearpag

%\clearpage
%\clearpag

\subsection{The nature of the third body}

If the outer star is a white dwarf or a main sequence star, its mass
is constrained to be $\lesssim0.5M_{\bigodot}$ by the lack of an
optical detection \citep{2001ApJ...563..934C}. The lack of absorbing
material along the line of sight to the X-ray source indicates that
the third star is not overflowing {\em its} Roche lobe. From the Roche
lobe fitting formula of \citep{1983ApJ...268..368E},
\be
R_3<R_{L}\approx{0.49q^{2/3}\over0.6q^{2/3}+\ln(1+q^{1/3})}a_{out},
\ee
where $q$ is the mass ratio of the third star to the total mass of the
inner binary. This translates to $R_3\lesssim0.36R_\odot$. From table
9 in \citet{2007ApJ...663..573B}, this implies
$m_3\lesssim0.39M_\odot$. We conclude that the only stars with mass
$\gtrsim0.4M_{\bigodot}$ that will fit into the outer orbit is a white
dwarf or neutron star (or black hole). This leaves open the
possibility that the third star is a main sequence star with
$m\lesssim0.4M_\odot$.

According to \citet{2008msah.conf..101I}, her Table 1, the fraction of
hierarchical triples consisting of a neutron star primary, a white
dwarf secondary, and a white dwarf tertiary formed via
binary\textemdash binary encounters is about $1.4\times10^{-3}$. The
fraction of triples consisting of a neutron star-white dwarf binary
orbited by a $0.4M_\odot$ (or lower) main sequence star is
similar. The fraction of neutron star-white dwarf-neutron star systems
is $2.1\times10^{-5}$. If this is the primary channel for formation of
triple star systems, the third star is likely to be either a white
dwarf or a low mass ($m<0.4M_\odot$) main sequence star.

\section{DISCUSSION} \label{sec:discussion}
The origin of the $170$ day luminosity variations in 4U 1820-30 was
first attributed to the presence of a third body in the system by
\citet{1988IAUS..126..347G}; this possibility was expanded upon by
\citet{2001ApJ...563..934C} and more recently by
\citet{2007MNRAS.377.1006Z}. \citet{2007MNRAS.377.1006Z} used a
numerical model that calculates the time evolution of an isolated
hierarchical triple of point masses, using secular perturbation theory
up to octupole terms. Their model neglects the effects of tidal and
rotational distortion of the white dwarf, tidal friction, mass
transfer and gravitational radiation from the inner binary. Their
calculations do include the GR periastron precession of the inner
binary.

\citet{2007MNRAS.377.1006Z} find a configuration that reproduces the
171 day oscillations (assuming they are due to variations in
$e$). They note that the GR precession rate is near $170$ days, and
then choose a rather low neutron star plus white dwarf mass of
$1.29+0.07M_\odot$. With this choice, the period of the GR precession
is $\sim168$ days. This period is very near, but slightly shorter
than, the observed 171 day period. To arrive at the longer period,
they choose the location and inclination of the third body so that the
Kozai torque results in a retrograde precession. When added to the GR
precession, this retrograde Kozai precession ensures the period of
eccentricity oscillations will be longer than 168 days. They are
driven to a much lower magnitude for the Kozai torque than employed in
this paper; they use $a_{out}=8.66a$ and $i_0=40.96^\circ$. They start
with $\omega=0^\circ$ and $e=10^{-4}$, ensuring that their solution
circulates rather than librating.

They note that the apparent near equality between the Kozai and GR
precession rates is ``a very remarkable coincidence'', but go on to
say that they do not have any explanation for this coincidence.

We have argued that the origin of the $171$ day period of the
luminosity variation of LMBX 4U 1820-30 arises from libration in the
Kozai resonance. This trapping explains why the Kozai precession rate
is comparable to the sum of the other precession rates in the
problem. If $k_2$ is small enough, then the largest precession
frequency in the absence of a third body is that given by general
relativity. In that case, the Kozai and GR precession rates will sum
to zero, i.e., the magnitude of the two precession rates will be
equal. Hence if $k_2$ is small, then the expansion of the orbit of the
inner binary naturally explains the ``remarkable coincidence'' noted
by \citet{2007MNRAS.377.1006Z}. We stress that, independent of the value
of $k_2$, the natural state of the system is likely to be libration
rather than circulation.

Trapping into libration is a consequence of mass-transfer driven
orbital expansion in the inner binary. We have pointed out that the
apparent negative period derivative, if it were intrinsic to the
system, would not last for reasonable fraction of the system's lifetime. We find this to be an
untenable situation.

The observed negative period derivative of the inner binary allows us
to constrain the tidal dissipation factor $Q$ yielding a very firm
lower limit of $(e/0.009)^2Q/k_2>3.15\times10^9$. We argue, however, that $(e/0.009)^2Q/k_2$ has to be yet higher, to trap and maintain the system in libration
around the stable Kozai fixed point. Our finding indicates that if 4U
1820-30 is indeed a triple system, the negative period derivative is
not an intrinsic property of the system. However, as we showed in section \ref{sec:constraints} it does not arise from the
acceleration of the gravitation field of the globular cluster in which
4U 1820-30 resides, as suggested by \citet{1993MNRAS.260..686V}.

%thermal tide discussion

In general, the eccentric orbit of a close binary system similar to 4U
1820-30 could lead to a time-dependent irradiation of the secondary
which could, in turn, give rise to a thermal tide
\citep{2010ApJ...714....1A}. A thermal tidal torque opposes the
gravitational tidal torque, tending to force the secondary away from
synchronous rotation and to enhance the orbital eccentricity. An
asynchronous spin may cause large tidal heating rates, depositing heat
in the interior of the secondary. In addition, the irradiation of the
stellar surface by the neutron star (or by the accretion disk) will
reduce the heat flux from the center of the white dwarf outward, so
these irradiated white dwarfs will be hotter than passively cooling
white dwarfs. Since they are hotter, they will have larger radii. The
interplay between the two tidal torques would eventually set the
equilibrium spin state. As long as this equilibrium state is not
reached, the resulting bulge may oscillate, causing a periodic
exchange of angular momentum between the orbit and the spin of the
white dwarf. This might provide an alternate mechanism for producing
the luminosity variations in 4U 1820-30. Since this period is very stable, $\dot
P_3/P_3<2.2\times10^{-4}$ according to \citet{2001ApJ...563..934C},  we are currently looking into possibility of such an interplay between gravitational and thermal tidal
torque as an explanation for $171$ day period in 4U 1820-30.

We anticipate that the resonance trapping mechanism we have described
in this paper is generic in Roche lobe overflow binaries in triple
systems. The exact nature of the librating orbit will vary with the
properties of the particular system. For example, for a binary with a
larger semimajor axis, such that $\dot\omega_{TB}>\dot\omega_{GR}$,
resonant trapping will lead to $|\dot\omega_{kozai}|=\dot\omega_{GR}$.

\section{CONCLUSIONS\label{sec:conclusions}}
This paper provides an estimate for a lower limit of the tidal dissipation
parameter $Q$ for a Helium white dwarf. It also elucidates the
possible evolutionary history of 4U 1820-30, i.e., how the system arrived
at a state where the secular dynamics are not dominated by the effects
of the white dwarf's tidal bulge, despite the fact that the white
dwarf is overflowing its Roche lobe in an orbit with a period of
$685\s$.

We suggest that the system is trapped in the Kozai resonance. This
resonance trapping is responsible for the observed $171$ day period,
which we interpret as the period of small oscillations around a stable
fixed point in the Kozai resonance. If the system is not librating,
one requires very fine tuning to get the 171 day period.

We provide lower limit on the tidal dissipation rate, as measured by
the factor $Q$; $(e/0.009)^2Q/k_2>4\times10^9$. 

Further exploration of the long term (tidal and mass overflow-driven)
evolution of this and similar short period ultra compact X-ray
binaries is clearly warranted. Inclusion of the thermal tides into dynamics of these systems may introduce an alternative explanation for origin of long period modulation of the light curve.
For the particular case of 4U 1820-30, better modelling of the
gravitational potential in the host globular cluster, NGC 6624, would
allow for an upper limit on $Q$. We are pursuing both lines
of investigation.

\acknowledgments The authors are grateful to Natasha Ivanova, Cole
Miller, Doug Hamilton, Andrew Cumming and Phil Arras for helpful discussions.  This
research has made use of the SIMBAD database, operated at CDS,
Strasbourg, France, and of NASA's Astrophysics Data System. 
The authors are supported in part by the Canada Research Chair program and
by NSERC of Canada.

%
%\appendix A
%
%
\vspace{15mm}
\section*{APPENDIX A\\
 EQUATIONS OF MOTION}

\noindent The equations of motion we employ model the Kozai interaction, the dynamical effects of the tidal bulge of the He white dwarf, GR periastron precession, the rotational bulge of the He white dwarf, conservative mass transfer driven by the emission of gravitational radiation,  and tidal dissipation. We do not consider tides raised on the neutron star primary. Detailed derivation of the equations representing Kozai cycles with tidal friction and GR periastron precession can be found in \citet{1998ApJ...499..853E} and \citet{2001ApJ...562.1012E}. Stellar masses are denoted by $m_1$ (the mass of the neutron star primary), $m_2$ (the mass of the white dwarf secondary) $M\equiv m_1+m_2$ (the inner binary mass), $m_3$ (the mass of the outer companion), and the reduced mass of the inner binary $\mu=m_{1}m_2/(m_{1}+m_2)$. The mean motion of the inner binary is $n=2\pi/P=[GM/a^{3}]^{1/2}$. The inner binary orbital elements are: semimajor axis $a$, eccentricity $e$, mutual inclination between the inner binary and the outer binary orbit $\imath$, the argument of periastron $\omega$, the longitude of ascending node $\Omega$. $k_{2}$ is the tidal Love number, $Q$ is the tidal dissipation factor, and $R_2$ is the radius of the white dwarf. The orbital parameters of the outer binary are denoted $a_{out}$ and $e_{out}$. $G$ is Newtons  constant and $c$ is the speed of light.

Changes in the semimajor axis of the inner binary $a$ are caused by tidal dissipation and mass transfer:
\be  \label{eq:a_dot} %$
      \dot{a}=\dot{a}_{TD}+\dot{a}_{MT},
\ee
where
\bea  \label{eq:a_dots}
      \dot{a}_{TD}&=&-2a\frac{1}{t_{F}}\left[\frac{1+\frac{15}{2}e^{2}+\frac{45}{8}e^{4}+\frac{5}{16}e^{6}}{(1-e^{2})^{\frac{13}{2}}}-\frac{\Omega_{h}}{n}\frac{1+3e^{2}+\frac{3}{8}e^{4}}{(1-e^{2})^{5}}\right]\label{eq:a_dot_TD}\\
&-&2a\frac{e^{2}}{1-e^{2}}\frac{9}{t_{F}}\left[\frac{1+\frac{15}{4}e^{2}+\frac{15}{8}e^{4}+\frac{5}{64}e^{6}}{(1-e^{2})^{\frac{13}{2}}}-\frac{11\Omega_{h}}{18n}\frac{1+\frac{3}{2}e^{2}+\frac{1}{8}e^{4}}{(1-e^{2})^{5}}\right]\\
     \dot{a}_{MT}&=&-\frac{2}{3} a\frac{\dot{m}_{2}}{m_2} 
\eea
with $\dot{m}_{2}$ given by:
\be \label{eq:m_dot}
  \dot{m}_{2}=-6.21\times 10^{-4}(\frac{m_{1}}{M_{\odot}})^{\frac{2}{3}}\left(\frac{P_{periastron}}{minutes}\right)^{\frac{-14}{3}} \frac{M_{\odot}}{\yr}.
 \ee
For the zero eccentricity case, equation \ref{eq:m_dot} is derived in detail in  \citet{1987ApJ...322..842R}, where instead of the dependancy on periastron period $P_{periastron}$ they consider dependancy on binary period.

The tidal friction time scale is
\be  \label{eq:tF}
t_{F}=\frac{1}{6}\left(\frac{a}{R_2}\right)^{5}\frac{1}{n}\frac{m_{2}}{m_{1}}\frac{Q}{k_{2}}.
\ee

The eccentricity of the inner binary $e$ is affected by the Kozai torque and by tidal dissipation:
\be \label{eq:e_dot}
    \dot{e}_{in}=\dot{e}_{Kozai}+\dot{e}_{TD},
\ee
where
\bea \label{eq:e_dots}
   \dot{e}_{Kozai}&=&\frac{15}{8}\frac{Gm_3}{a_{out}^3(1-e^{2}_{out})^{\frac{3}{2}}n}e\sqrt{1-e^{2}}\sin2\omega\sin^{2}\imath \\
   \dot{e}_{TD}&=& -\frac{9e}{t_{F}}\left[\frac{1+\frac{15}{4}e^{2}+\frac{15}{8}e^{4}+\frac{5}{64}e^{6}}{(1-e^{2})^{\frac{13}{2}}}-\frac{11\Omega_{h}}{18n}\frac{1+\frac{3}{2}e^{2}+\frac{1}{8}e^{4}}{(1-e^{2})^{5}}\right]\label{eq:e_dot_TD}.\\
 \eea 

The mutual inclination between the inner and the outer binary orbit, $\imath$ is affected by Kozai torques, by the rotational bulge, and by tidal dissipation:
\be  \label{eq:i_dot}
\dot{\imath}=\dot{\imath}_{Kozai}+\dot{\imath}_{RB}+\dot{\imath}_{TD},
\ee
where
\bea \label{eq:i_dots}
\dot{\imath}_{Kozai}&=&-\frac{15}{8}\frac{Gm_3}{a_{out}^3(1-e^{2}_{out})^{\frac{3}{2}}n}\frac{e^{2}}{\sqrt{1-e^{2}}}\sin2\omega\sin\imath\cos\imath \\
\dot{\imath}_{RB}&=&\frac{m_{1}k_{2}R_2^{5}}{2\mu na^{5}}\frac{\Omega_{h}(\Omega_{q}\sin\omega-\Omega_{e}\cos\omega)}{(1-e^2)^5}\\
\dot{\imath}_{TD}&=&-\frac{\Omega_{e}\sin\omega}{2nt_{F}}\frac{1+\frac{3}{2}e^{2}+\frac{1}{8}e^{4}}{(1-e^{2})^{5}}-\frac{\Omega_{q}\cos\omega}{2nt_{F}}\frac{1+\frac{9}{2}e^{2}+\frac{5}{8}e^{4}}{(1-e^{2})^{5}}.
\eea

Besides the negative precession rate of the argument of periastron due to Kozai cycles, the total precession rate of the argument of periastron has additional positive contributions from the tidal bulge, GR, the rotational bulge, and the tidal dissipation:
\be \label{eq: omega_dot}
\dot{\omega}_{in}=\dot{\omega}_{Kozai}+\dot{\omega}_{TB}+\dot{\omega}_{GR}+\dot{\omega}_{RB}+\dot{\omega}_{TD},
\ee
\noindent where
\bea \label{eq:omega_dots}
\dot{\omega}_{Kozai}&=&\frac{3}{4}\frac{Gm_3}{a_{out}^3(1-e^{2}_{out})^{\frac{3}{2}}n}\frac{1}{\sqrt{1-e^{2}}}\Bigg[ 2(1-e^{2})+5\sin^{2}\omega(e^{2}-\sin^{2}\imath)\Bigg] \\
\dot{\omega}_{TB}&=&\frac{15(GM)^{\frac{1}{2}}}{16a^{\frac{13}{2}}}\frac{8+12e^{2}+e^{4}}{(1-e^{2})^{5}}\frac{m_{1}}{m_{2}}k_{2}R_2^{5}\\
\dot{\omega}_{GR}&=&\frac{3(GM)^{\frac{3}{2}}}{a^{\frac{5}{2}}c^{2}(1-e^{2})}\\
\dot{\omega}_{RB}&=&\frac{M^{\frac{1}{2}}}{4G^{\frac{1}{2}}a^{\frac{7}{2}}(1-e^{2})^{2}}\frac{k_{2}R_2^{5}}{m_{2}}\\
 &\times&\Bigg[\left(2\Omega^{2}_{h}-\Omega^{2}_{e}-\Omega^{2}_{q}\right)+2\Omega_{h}\cot\imath\left(\Omega_{e}\sin\omega +\Omega_{q}\cos\omega\right)\Bigg]\\
\dot{\omega}_{TD}&=&-\frac{\Omega_e\cos(\omega)\cot\imath}{2nt_F}\frac{1+\frac{3}{2}e^2+\frac{1}{8}e^4}{(1-e^2)^5}+\frac{\Omega_q\sin(\omega)\cot\imath}{2nt_F}\frac{1+\frac{9}{2}e^2+\frac{5}{8}e^4}{(1-e^2)^5}.
\eea
The precession of the longitude of ascending node is caused by Kozai cycles, rotational bulge and tidal dissipation:
\be  \label{eq:Omega_dot}
\dot{\Omega}_{in}=\dot{\Omega}_{Kozai}+\dot{\Omega}_{RB}+\dot{\Omega}_{TD},
\ee
where
\bea \label{eq:Omega_dots}
\dot{\Omega}_{Kozai}&=&-\frac{Gm_3}{a_{out}^3(1-e^{2}_{out})^{\frac{3}{2}}n}\frac{\cos\iota}{4\sqrt{1-e^{2}}}\Bigg(3+12e^{2}-15e^{2}\cos^{2}\omega\Bigg)\\
\dot{\Omega}_{RB}&=&\frac{m_{1}k_{2}R_2^{5}}{2\mu na^{5}\sin\imath}\frac{\Omega_{h}}{(1-e^{2})^{2}}\Bigg(-\Omega_{q}\cos\omega-\Omega_{e}\sin\omega\Bigg)\\
\dot{\Omega}_{TD}&=&\frac{\Omega_{e}\cos\omega}{2n\sin\imath t_{F}}\frac{1+\frac{3}{2}e^{2}+\frac{1}{8}e^{4}}{(1-e^{2})^{5}}-\frac{\Omega_{q}\sin\omega}{2n\sin\imath t_{F}}\frac{1+\frac{9}{2}e^{2}+\frac{5}{8}e^{4}}{(1-e^{2})^{5}}.
\eea

%
%\appendix B
%
%
\vspace{20mm}
\section*{APPENDIX B\\
 ADIABATIC INVARIANCE OF THE ACTION}
 
 \noindent Time-dependent Hamiltonians, even those with just one degree of freedom, can be difficult to solve. However, for Hamiltonians where the time dependance is sufficiently slow, the problem is easier to tackle due to the existence of variables that are almost constant. The approximate constants are the action variables of the Hamiltonian, when the slow time dependance is neglected. Suppose that the time dependance enters through a time dependent parameter $\kappa(t)$. If the parameter $\kappa$ varies very slowly with time, treating $\kappa$ as time-independent parameter allows us to find action-angle variables following the standard prescription. These action-angle variables are function of time through $\kappa(t)$, which leads to the action no longer being a constant of motion. However, when $\kappa$ varies slowly with time, the action is nearly constant. Such an action is known as an adiabatic invariant. 
 
 As described in section \ref{sec:capture}, capture in the resonance is a natural consequence of semimajor axis expansion driven by mass transfer from the white dwarf. The Hamiltonian of our system (see equation \ref{eq:hamiltonian}) is a function of the semimajor axis, which is a parameter of $H$, playing the role of $\kappa(t)$. In our case the semimajor axis is not the only parameter varying with time; the masses of the inner binary vary with time as well.  Here we show, both analytically and via numerical integration, that the change in the eccentricity is coupled to the change in the semimajor axis. When the semimajor axis expands (respectively, contracts) the eccentricity of the stable fixed point increases (decreases). We also demonstrate that the timescale for the change in the eccentricity is a factor of $\gtrsim 150$ shorter than the timescale for the semimajor axis.   
 
 To find the action we expand our Hamiltonian (equation \ref{eq:hamiltonian}) around the fixed point:
 \bea
 \G=\G_f + \Delta \G\\
 \omega=\omega_f +\Delta \omega
 \eea
 
 \noindent Since we are expanding around the resonance, all terms $\propto \Delta \G$ vanish. After some algebra we find:
 
 \bea \label{eq:exp_hamil}
 H&=&-A\Big[-10+12\frac{\G_f^2} {\La^2} \cos^2{\imath_f}+9\frac{\G_f^2}{\La^2}+15\cos^2{\imath_f}+\frac{B}{A} \frac{\La}{\G_f}+k_2\frac{C}{A}\left( 35\frac{\La^9}{\G_f^9}-30\frac{\La^7}{\G_f^7}+3\frac{\La^5}{\G_f^5}\right)+k_2\frac{D}{A} \frac{\La^3}{\G_f^3}\Big]\nonumber\\&&-\frac{A}{2\La^2}\Big[18-24\cos^2{\imath_f}+2\frac{B}{A}\frac{\La^3}{\G_f^3}+30k_2\frac{C}{A}\left(105\frac{\La^11}{\G_f^11}-56\frac{\La^9}{G_f^9}+3\frac{\La^7}{G_f^7}\right)+12k_2\frac{D}{A}\frac{\La^5}{G_f^5}\Big]\Delta\G^2\nonumber\\&&-15A\left(1-\frac{\G_f^2}{\La^2}\right)\sin^2{\imath_f}\Delta\omega^2
  \eea
  \noindent which is similar to the Hamiltonian of the harmonic oscillator. Written more compactly (and implicitly defining $\alpha(t)$, $\beta(t)$ and $\C(t)$):
  \be\label{eq:hamil_for_ho}
  H= \C(t)+\frac{\alpha(t)}{2}\Delta\G^2+\frac{\beta(t)}{2}\Delta\omega^2=H_0.
  \ee
  \noindent We solve for $\Delta\G$ and evaluate the integral
  \be
  J=\frac{2}{\pi}\int_0^{\Delta\omega_{max}}\left(\frac{2(H_0-\C(t))}{\alpha}-\frac{\beta}{\alpha}\right)^{\frac{1}{2}}\,d \Delta\omega
  \ee
  \noindent where $\Delta\omega_{max}=\left(2(H_0-\C(t))/\beta\right)^{\frac{1}{2}}$. We find:
  \be\label{eq:action_ho}
  J=\frac{H_0-\C(t)}{(\alpha\beta)^{\frac{1}{2}}}.
  \ee
  \noindent Plugging in the corresponding terms from equation (\ref{eq:exp_hamil}) yields:
  
  \be
  J=\frac{\La(a,m_1,m_2)}{e_f\sin{\imath_f}}\! \, \frac{P_1(e_f, a, m_1, m_2, \imath_f)}{P_2^\frac{1}{2}(e_f, a, m_1, m_2, \imath_f)},
  \ee
  \noindent where
  \bea
  P_1&=&\frac{H_0}{A}-10-12(1-e_f^2)\cos^2{\imath_f}+9(1-e_f^2)+15\cos^2{\imath_f}+\frac{B}{A}(1+\frac{1}{2}e_f^2)+4k_2\frac{C}{A}(2+15e_f^2)\nonumber\\&&+k_2\frac{D}{A}(1+\frac{3}{2}e_f^2)\\
  \eea
  and
  \be
  P_2=30\Big[18-24\cos^2{\imath_f}+2\frac{B}{A}(1+\frac{3}{2}e_f^2)+120k_2\frac{C}{A}(13+84e_f^2)+12k_2\frac{D}{A}(1+\frac{5}{2}e_f^2).
  \ee
  Since the action $J$ is an adiabatic invariant, we have:
  \be\label{eq:dJdt}
  \frac{dJ}{dt}=\frac{\partial J}{\partial e}\dot{e}+\frac{\partial J}{\partial a}\dot{a}+\left(\frac{\partial J}{\partial m_2}-\frac{\partial J}{\partial m_1}\right)\dot{m}_2=0.
  \ee
  The partial derivatives are:
  \bea
  \frac{\partial J}{\partial e}&=&-\frac{J}{e_f}\left(1-\frac{e_f}{P_1}\frac{\partial P_1}{\partial e}+\frac{1}{2}\frac{e_f}{P_2}\frac{\partial P_2}{\partial e}\right)=\C_e\frac{J}{e_f}\\
  \frac{\partial J}{\partial a}&=&\frac{J}{a}\left(1+\frac{a}{P_1}\frac{\partial P_1}{\partial a}-\frac{1}{2}\frac{a}{P_2}\frac{\partial P_2}{\partial a}\right)=\C_a\frac{J}{a}\\
  \frac{\partial J}{\partial m_2}&=&\frac{J}{m_2}\left(1+\frac{m_2}{P_1}\frac{\partial P_1}{\partial m_2}-\frac{1}{2}\frac{m_2}{P_2}\frac{\partial P_2}{\partial m_2}\right)=\C_{m_2}\frac{J}{m_2}\\
  \frac{\partial J}{\partial m_1}&=&\frac{J}{m_1}\left(1+\frac{m_1}{P_1}\frac{\partial P_1}{\partial m_1}-\frac{1}{2}\frac{m_1}{P_2}\frac{\partial P_2}{\partial m_1}\right)=\C_{m_1}\frac{J}{m_1}.
  \eea
  Plugging these partial derivatives back into equation (\ref{eq:dJdt}) yields:
  \be\label{eq:all_dots}
  0=\C_e\frac{\dot{e}}{e_f}+\C_a\frac{\dot{a}}{a}+\left(\C_{m_2}-\frac{m_2}{m_1}\C_{m_1}\right)\frac{\dot{m}_2}{m_2}.
  \ee
  
  The inner binary orbit is eccentric, which makes the mass transfer rate proportional to periastron distance $r_p=a(1-e)$. Hence the $\dot{m_2}$ term can be decoupled into two terms, one proportional to $\dot e$ and the other proportional to $\dot a$:
  \be\label{eq:m2decouple}
 \frac{\dot{m}_2}{m_2}=\frac{\dot{r_p}}{r_p}=-\frac{3}{2}\frac{\dot a}{a}+\frac{3}{2}\frac{\dot e}{1-e}.
  \ee 
  Combining equations (\ref{eq:all_dots}) and (\ref{eq:m2decouple}) and solving for $\dot e$ leads to:
  \be
  \frac{\dot e}{e_f}=\frac{\frac{3}{2}\left(\C_{m_2}-\frac{m_2}{m_1}\C_{m_1}\right)-\C_a}{\C_e+\frac{3}{2}\frac{e_f}{1-e_f}\left(\C_{m_2}-\frac{m_2}{m_1}\C_{m_1}\right)}\frac{\dot a}{a}.
  \ee
  Plugging in the numerical values:
  \be\label{eq:edot_vs_adot}
  \frac{\dot e}{e_f}\approx150\frac{\dot a}{a}.
  \ee
  Defining the time scales for the eccentricity and the semimajor axis  to decay or increase (depending on the value of $Q$) as $\tau_e=e_f/\dot e$ and $\tau_a=a/\dot a$, the timescales in equation (\ref{eq:edot_vs_adot}) are related by:
  \be
  \tau_e\sim6.7\times10^{-3}\tau_a.
  \ee
  
  %\clearpage
\begin{figure}
\epsscale{0.55} 
\plotone{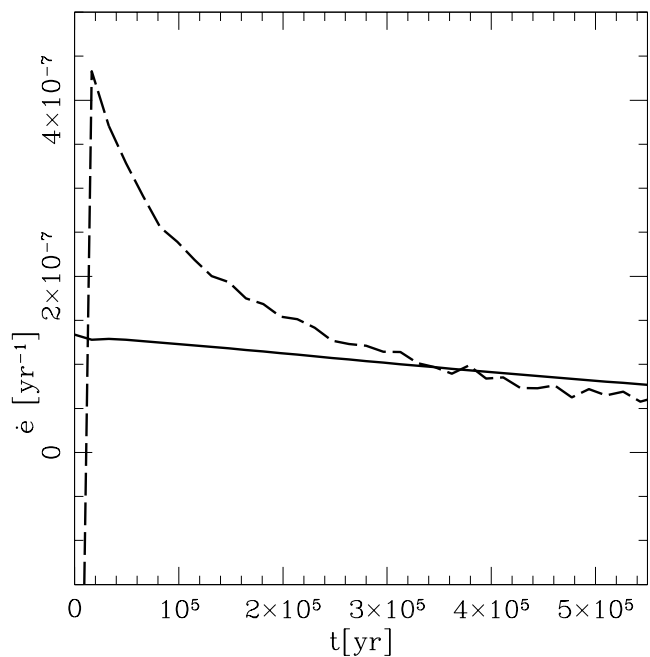}
\caption{
 $\dot e$ as a function of time for the case where the semimajor axis is expanding, $Q=8\times10^7$. We start integration exactly at the fixed point, where initial eccentricity is $e_{f,0}=0.01555$. All other parameters are as listed in table 1. The solid line comes from the analytic estimate where the action $J$ is considered to be an adiabatic invariant. The dashed line is a result of numerical integration. As expected from the action being adiabatic invariant, $\dot e$ is positive. The difference in the magnitude of $\dot e$ within first $2\times10^5\yr$ is a result of our simplified analytic calculation that does not include spin dynamics. Since our analytic estimate is valid for small eccentricities, here we stop the integration when $e>0.1$   \label{Fig:e_dot8Q7}}
\end{figure}
%\clearpage
  
  To demonstrate that the eccentricity evolution is indeed a consequence of the action being an adiabatic invariant, we follow the evolution of the orbit around the fixed point $e_f=0.0155$ and $\omega_f=90^\circ$. Figure \ref{Fig:e_dot8Q7} shows $\dot e$ as a function of time in a case where the semimajor axis is increasing, meaning that tidal dissipation is sufficiently weak so that the evolution of the semimajor axis is dominated by mass transfer ($Q=8\times10^7$). The solid line presents $\dot e$ predicted by equation (\ref{eq:edot_vs_adot}). For $t\gtrsim10^{5}\yr$ the numerical integration gives $\dot e \approx 10^{-7}\yr^{-1}$ corresponding to a timescale $150$ times shorter than the timescale for the semimajor axis. 
  
  Despite the fact that the semimajor axis is expanding, the numerical integration shows a transient phase (roughly the first 2000 years) where $de/dt<0$, and a longer phase ($\sim 10^5\yr$) where $\dot e$ is larger than predicted by equation (\ref{eq:edot_vs_adot}). There are contributions to the eccentricity evolution which we have ignored in our analytic treatment; for example the spin of the white dwarf is not locked during the evolution of the system. These un-modelled contributions are the source of the transient behaviour. 
  
\begin{figure}
\epsscale{0.55} 
\plotone{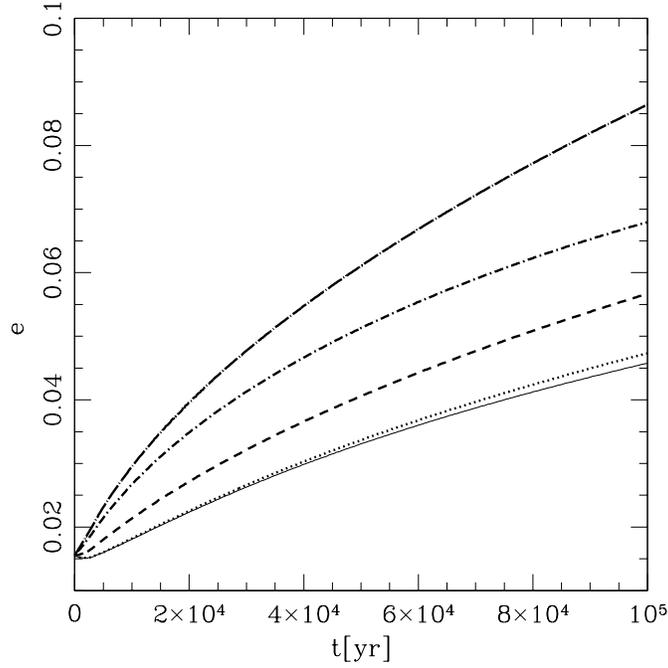}
\caption{
 The eccentricity of the fixed point of the inner binary as a function of time. Initial conditions are the same as in Figure \ref{Fig:e_dot8Q7}. The solid line is a result of the numerical integration including all dynamical effects, while the dotted line is a result of the same integration but with $\dot e_{TD}=0$; these two results demonstrate that direct tidal dissipation on the eccentricity does not significantly affect the evolution of eccentricity. The dashed line is the result of integration where $\dot a_{TD}=0$; this term has a more significant effect on the evolution of the eccentricity. The dash-dotted line presents the case where we set $Q=\infty$, but only in the equations that govern the spin evolution.  The long dash-dotted line shows the eccentricity evolution when $Q=\infty$ for the spins and $\dot a_{TD}=0$; in this case we have the fastest increase in the eccentricity.  \label{Fig:e_all8Q7}}
\end{figure}

 To support this statement, we illustrate the eccentricity evolution in various  cases where we turn off different dynamical effects in Figure \ref{Fig:e_all8Q7}. The solid line presents a result from the numerical integration that includes all dynamical effects in our model, while the dotted line is the same integration with the $\dot e_{TD}$ term set to $0$; the result shows that direct tidal dissipation on the eccentricity (equation \ref{eq:e_dot_TD}) is not dynamically significant. The dashed line presents the case where $\dot a_{TD}$ is set to $0$ (see equation \ref{eq:a_dot_TD});  The result shows that $\dot a_{TD}$ has a significant influence on the eccentricity evolution. The dash-dotted line shows the eccentricity when tidal dissipation factor $Q$ is set to infinity, but only in the differential equations that govern spin evolution. The long dash-dotted line shows the eccentricity evolution in the case where $Q$ is set to infinity in the equations that govern the evolution of the spins together with $\dot a_{TD}$ being set to $0$. The latter three cases demonstrate that the eccentricity starts increasing immediately with the semimajor axis expansion, which is exactly the behaviour predicted by the analytic analysis.  After $5.5\times10^5\yr$ the eccentricity becomes $\gtrsim0.1$ and since our analytic estimate is valid only for small eccentricities we stop the integration here. 
    
  The cause of the transient behaviour is the fact that the spin of the white dwarf is not, contrary to our choice of initial conditions, tidally locked during the evolution of the system. Whether the spin settles down in some Cassini state or other stable configuration later during the evolution of the system is a possibility open to further investigation.

%\clearpage
\clearpage

\bibliography{1820}{}

\end{document}

%% file: table1.tex
%\documentclass{aastex}
%\bibliographystyle{apj}
%\begin{document}
%\newcommand       \cm		{\,{\rm cm }}

\begin{table}
\begin{center}
\begin{tabular}{|l|l|l|l|}
\tableline
\multicolumn{4}{|c|}{TABLE 1. System parameters} \\
\tableline
\tableline
Symbol & Definition & Value & Reference \\ \hline
$m_1$ & Neutron star (primary) mass & $1.4 M_{\bigodot}$ &\\
$m_2$ & White dwarf (secondary) mass & $0.067 M_{\bigodot}$ &\citet{1987ApJ...322..842R}\\
$m_3$ & Third companion mass & $0.55 M_{\bigodot}$& \\
$a_1$ & Inner binary semimajor axis & $1.32\times10^{10}\cm$ & \citet{1987ApJ...312L..17S}\\
$a_{out}$ & Outer binary semimajor axis & $8.0a_1$ &\\
$e_{in, 0}$ & Inner binary initial eccentricity & $0.009$ &\\
$e_{out, 0}$ & Outer binary eccentricity & $10^{-4}$ &\\
$i_{init}$ & Initial mutual inclination & $44.715^{o}$  & \\
$\omega_{in, 0}$ & Initial argumet of periastron & $90^{o}$ &\\
$\Omega_{in}$ & Longitude of ascending node & $0$ &\\
$R_2$ & White dwarf radius & $2.2\times10^9 \cm$ &\\
$k_2$ & Tidal Love number & $0.01$ & Arras (private communication)\\
$Q$ & Tidal dissipation factor & $5\times10^7$ &\\
\tableline
\tableline
\end{tabular}
\end{center}
\end{table}

%\bibliography{1820}{}

%\end{document}

%% file: article_last_arxiv.bbl
\begin{thebibliography}{26}
\expandafter\ifx\csname natexlab\endcsname\relax\def\natexlab#1{#1}\fi

\bibitem[{{Anderson} {et~al.}(1997){Anderson}, {Margon}, {Deutsch}, {Downes},
  \& {Allen}}]{1997ApJ...482L..69A}
{Anderson}, S.~F., {Margon}, B., {Deutsch}, E.~W., {Downes}, R.~A., \& {Allen},
  R.~G. 1997, \apjl, 482, L69+

\bibitem[{{Arras} \& {Socrates}(2010)}]{2010ApJ...714....1A}
{Arras}, P., \& {Socrates}, A. 2010, \apj, 714, 1

\bibitem[{{Beatty} {et~al.}(2007){Beatty}, {Fern{\'a}ndez}, {Latham}, {Bakos},
  {Kov{\'a}cs}, {Noyes}, {Stefanik}, {Torres}, {Everett}, \&
  {Hergenrother}}]{2007ApJ...663..573B}
{Beatty}, T.~G., {et~al.} 2007, \apj, 663, 573

\bibitem[{{Chandrasekhar}(1939)}]{1939isss.book.....C}
{Chandrasekhar}, S. 1939, {An introduction to the study of stellar structure}

\bibitem[{{Chou} \& {Grindlay}(2001)}]{2001ApJ...563..934C}
{Chou}, Y., \& {Grindlay}, J.~E. 2001, \apj, 563, 934

\bibitem[{{Eggleton}(1983)}]{1983ApJ...268..368E}
{Eggleton}, P.~P. 1983, \apj, 268, 368

\bibitem[{{Eggleton} {et~al.}(1998){Eggleton}, {Kiseleva}, \&
  {Hut}}]{1998ApJ...499..853E}
{Eggleton}, P.~P., {Kiseleva}, L.~G., \& {Hut}, P. 1998, \apj, 499, 853

\bibitem[{{Eggleton} \& {Kiseleva-Eggleton}(2001)}]{2001ApJ...562.1012E}
{Eggleton}, P.~P., \& {Kiseleva-Eggleton}, L. 2001, \apj, 562, 1012

\bibitem[{{Fabrycky} \& {Tremaine}(2007)}]{2007ApJ...669.1298F}
{Fabrycky}, D., \& {Tremaine}, S. 2007, \apj, 669, 1298

\bibitem[{{Ford} {et~al.}(2000){Ford}, {Kozinsky}, \&
  {Rasio}}]{2000ApJ...535..385F}
{Ford}, E.~B., {Kozinsky}, B., \& {Rasio}, F.~A. 2000, \apj, 535, 385

\bibitem[{{Grindlay}(1988)}]{1988IAUS..126..347G}
{Grindlay}, J.~E. 1988, in IAU Symposium, Vol. 126, The Harlow-Shapley
  Symposium on Globular Cluster Systems in Galaxies, ed. {J.~E.~Grindlay \&
  A.~G.~D.~Philip}, 347--363

\bibitem[{{Innanen} {et~al.}(1997){Innanen}, {Zheng}, {Mikkola}, \&
  {Valtonen}}]{1997AJ....113.1915I}
{Innanen}, K.~A., {Zheng}, J.~Q., {Mikkola}, S., \& {Valtonen}, M.~J. 1997,
  \aj, 113, 1915

\bibitem[{{Ivanova}(2008)}]{2008msah.conf..101I}
{Ivanova}, N. 2008, in Multiple Stars Across the H-R Diagram, ed. {S.~Hubrig,
  M.~Petr-Gotzens, \& A.~Tokovinin}, 101--+

\bibitem[{{King} {et~al.}(1993){King}, {Stanford}, {Albrecht}, {Barbieri},
  {Blades}, {Boksenberg}, {Crane}, {Disney}, {Deharveng}, {Jakobsen},
  {Kamperman}, {Macchetto}, {Mackay}, {Paresce}, {Weigelt}, {Baxter},
  {Greenfield}, {Jedrzejewski}, {Nota}, {Sparks}, \&
  {Sosin}}]{1993ApJ...413L.117K}
{King}, I.~R., {et~al.} 1993, \apjl, 413, L117

\bibitem[{{Kozai}(1962)}]{1962AJ.....67..591K}
{Kozai}, Y. 1962, \aj, 67, 591

\bibitem[{{Kuulkers} {et~al.}(2003){Kuulkers}, {den Hartog}, {in't Zand},
  {Verbunt}, {Harris}, \& {Cocchi}}]{2003A&A...399..663K}
{Kuulkers}, E., {den Hartog}, P.~R., {in't Zand}, J.~J.~M., {Verbunt},
  F.~W.~M., {Harris}, W.~E., \& {Cocchi}, M. 2003, \aap, 399, 663

\bibitem[{{Larwood}(1998)}]{1998MNRAS.299L..32L}
{Larwood}, J. 1998, \mnras, 299, L32+

\bibitem[{{Rappaport} {et~al.}(1982){Rappaport}, {Joss}, \&
  {Webbink}}]{1982ApJ...254..616R}
{Rappaport}, S., {Joss}, P.~C., \& {Webbink}, R.~F. 1982, \apj, 254, 616

\bibitem[{{Rappaport} {et~al.}(1987){Rappaport}, {Ma}, {Joss}, \&
  {Nelson}}]{1987ApJ...322..842R}
{Rappaport}, S., {Ma}, C.~P., {Joss}, P.~C., \& {Nelson}, L.~A. 1987, \apj,
  322, 842

\bibitem[{{Reg{\"o}s} {et~al.}(2005){Reg{\"o}s}, {Bailey}, \&
  {Mardling}}]{2005MNRAS.358..544R}
{Reg{\"o}s}, E., {Bailey}, V.~C., \& {Mardling}, R. 2005, \mnras, 358, 544

\bibitem[{{Stella} {et~al.}(1987){Stella}, {Priedhorsky}, \&
  {White}}]{1987ApJ...312L..17S}
{Stella}, L., {Priedhorsky}, W., \& {White}, N.~E. 1987, \apjl, 312, L17

\bibitem[{{van der Klis} {et~al.}(1993{\natexlab{a}}){van der Klis},
  {Hasinger}, {Verbunt}, {van Paradijs}, {Belloni}, \&
  {Lewin}}]{1993A&A...279L..21V}
{van der Klis}, M., {Hasinger}, G., {Verbunt}, F., {van Paradijs}, J.,
  {Belloni}, T., \& {Lewin}, W.~H.~G. 1993{\natexlab{a}}, \aap, 279, L21

\bibitem[{{van der Klis} {et~al.}(1993{\natexlab{b}}){van der Klis},
  {Hasinger}, {Dotani}, {Mitsuda}, {Verbunt}, {Murphy}, {van Paradijs},
  {Belloni}, {Makishina}, {Morgan}, \& {Lewin}}]{1993MNRAS.260..686V}
{van der Klis}, M., {et~al.} 1993{\natexlab{b}}, \mnras, 260, 686

\bibitem[{{Wijers} \& {Pringle}(1999)}]{1999MNRAS.308..207W}
{Wijers}, R.~A.~M.~J., \& {Pringle}, J.~E. 1999, \mnras, 308, 207

\bibitem[{{Wu} {et~al.}(2007){Wu}, {Murray}, \&
  {Ramsahai}}]{2007ApJ...670..820W}
{Wu}, Y., {Murray}, N.~W., \& {Ramsahai}, J.~M. 2007, \apj, 670, 820

\bibitem[{{Zdziarski} {et~al.}(2007){Zdziarski}, {Wen}, \&
  {Gierli{\'n}ski}}]{2007MNRAS.377.1006Z}
{Zdziarski}, A.~A., {Wen}, L., \& {Gierli{\'n}ski}, M. 2007, \mnras, 377, 1006

\end{thebibliography}
